\documentclass[twocolumn,epjc3]{svjour3}
\usepackage{graphics}

\usepackage{cuted}
\usepackage{flushend}

\usepackage{microtype}
\usepackage{bbm}
\usepackage{amsmath}
\usepackage{mathtools}
\usepackage{amssymb}
\usepackage{mathrsfs}       
\usepackage[pdftex]{graphicx}
\usepackage{pgfplots}
\pgfplotsset{width=10cm,compat=1.9}
\usepackage{xargs}       
\usepackage{wasysym}
\usepackage{engrec}
\usepackage{enumerate}
\usepackage{appendix}
\usepackage{empheq}
\usepackage{float}
\usepackage{stmaryrd}
\usepackage[parfill]{parskip}
\usepackage[pdftex,breaklinks]{hyperref}
\usepackage{titletoc}
\usepackage{makecell}
\usepackage{lscape}
\usepackage{algorithm}
\usepackage{algorithmicx}
\usepackage{tensor}
\usepackage{algpseudocode}
\usepackage{blkarray}
\usepackage{multirow}
\usepackage{bigstrut}
\usetikzlibrary{plotmarks}
\usepackage{braket}

\usepackage{simpler-wick}
\usepackage{bm}

\begin{document}
\title{\textit{Ab initio} description of monopole resonances in light- and medium-mass nuclei}
\subtitle{I. Technical aspects and uncertainties of \textit{ab initio} PGCM calculations}
\author{A. Porro\thanksref{ad:tud,ad:emmi,ad:saclay} 
\and T. Duguet\thanksref{ad:saclay,ad:kul}
\and J.-P. Ebran\thanksref{ad:dam,ad:dam_u}
\and M. Frosini\thanksref{ad:cadarache}
\and R. Roth\thanksref{ad:tud,ad:darm2}
\and V. Som\`a\thanksref{ad:saclay}}

\institute{
\label{ad:tud}
Technische Universit\"at Darmstadt, Department of Physics, 64289 Darmstadt, Germany
\and
\label{ad:emmi}
ExtreMe Matter Institute EMMI, GSI Helmholtzzentrum f\"ur Schwerionenforschung GmbH, 64291 Darmstadt, Germany
\and
\label{ad:saclay}
IRFU, CEA, Universit\'e Paris-Saclay, 91191 Gif-sur-Yvette, France 
\and
\label{ad:kul}
KU Leuven, Department of Physics and Astronomy, Instituut voor Kern- en Stralingsfysica, 3001 Leuven, Belgium
\and
\label{ad:dam}
CEA, DAM, DIF, 91297 Arpajon, France
\and
\label{ad:dam_u}
Universit\'e Paris-Saclay, CEA, Laboratoire Mati\`ere en Conditions Extr\^emes, 91680 Bruy\`eres-le-Ch\^atel, France
\and
\label{ad:cadarache}
CEA, DES, IRESNE, DER, SPRC, 13108 Saint-Paul-l\`es-Durance, France
\and
\label{ad:darm2}
Helmholtz Forschungsakademie Hessen f\"ur FAIR, GSI Helmholtzzentrum, 64289 Darmstadt, Germany
}

\date{Received: \today{} / Revised version: date}

\maketitle
%
%
\begin{abstract}
Giant resonances (GRs) are a striking manifestation of collective motions in mesoscopic systems such as atomic nuclei. Until recently, theoretical investigations have essentially relied on the (quasiparticle) random phase approximation ((Q)RPA), and extensions of it, based on phenomenological energy density functionals (EDFs). As part of a current effort to describe GRs within an \textit{ab initio} theoretical scheme, the present work promotes the use of the projected generator coordinate method (PGCM). This method, which can handle anharmonic effects while satisfying symmetries of the nuclear Hamiltonian, displays a favorable (i.e. mean-field-like) scaling with system's size. Presently focusing on the isoscalar giant monopole resonance (GMR) of light- and medium-mass nuclei, PGCM's potential to deliver wide-range \textit{ab initio} studies of GRs in closed- and open-shell nuclei encompassing pairing, deformation, and shape coexistence effects is demonstrated. The comparison with consistent QRPA calculations highlights PGCM's unique attributes and sheds light on the intricate interplay of nuclear collective excitations. The present paper is the first in a series of four and focuses on technical aspects and uncertainty quantification of \textit{ab initio} PGCM calculations of GMR using the doubly open-shell $^{46}$Ti as an illustrative example. The second paper displays results for a set of nuclei of physical interest and proceeds to the comparison with consistent (deformed) \textit{ab initio} QRPA calculations. While the third paper analyzes useful moments of the monopolar strength function and different ways to access them within PGCM calculations, the fourth paper focuses on the effect of the symmetry restoration on the monopole strength function.
\end{abstract}

\section{Introduction}

Giant resonances designate specific excitations of the system in which a significant fraction of the nucleons are involved in the process. Giant resonances are categorised according to their multipolarity and isospin nature, i.e. isoscalar or isovector, and are best pictured in terms of vibrations of the nuclear surface in a liquid-drop approach. The isoscalar GMR addressed in this work, also referred to as the \textit{breathing mode}, involves $J^\pi = 0^+$ excitations in which neutrons and protons oscillate radially in phase. As such, the GMR provides valuable information about the incompressibility of infinite nuclear matter~\cite{Blaizot1976,Blaizot80a,Khan13a,Garg2018a}, a key quantity characterising the nuclear equation of state (EoS).

While GRs are to a large extent made out of a coherent sum of 1-particle/1-hole excitations, their coupling to the background of many-particle/many-hole (ph) excitations relates to a well known phenomenon in solid state physics and is referred to as Landau damping. Furthermore, GRs lie above the particle emission threshold. Consequently, they couple to the particle  continuum, thus inducing a so-called escape width. Last but not least, they can also be damped through coupling to the electromagnetic field leading to the emission of a photon~\cite{Bertsch83a,Bortignon98a}.

The theoretical description of GRs is a mature field. The fact that they are dominantly made up of a coherent sum of 1-particle/1-hole excitations is at the heart of the usefulness of (Q)RPA and associated extensions based on phenomenological EDFs. In the last 15 years, systematic studies over the entire nuclear chart via full-fledged deformed QRPA calculations have become doable~\cite{Peru07a,Yoshida08a,Peru08a,Losa10a,Martini12a,Peru19a}. Regarding the GMR in particular, calculations have led to an improved understanding of several features such as the effect of its coupling to the Giant Quadrupole Resonance (GQR) in doubly-open shell nuclei~\cite{Peru08a} that is addressed in the present work. However, certain aspects are left unexplained so far, e.g. the evolution of the GMR centroid along semi-magic isotopic chains away from doubly closed-shell nuclei~\cite{Khan09a,Khan09b}. At the same time, beyond-RPA calculations such as self-consistent-RPA (SCRPA) (see Ref.~\cite{Schuck21a} for a recent review), Second RPA (SRPA)~\cite{Gambacurta10a,Gambacurta16a} or particle-vibration coupling (PVC)~\cite{Bortignon98a,Litvinova07a,Li22a} techniques have shown to be instrumental to quantitatively address refined aspects of GRs such as their (Landau) width.

Another class of theories relates to the time-dependent description of nuclear systems following the application of an external perturbation. If the time evolution is performed on a Slater determinant (Bogoliubov vacuum) such a theory can be shown to be strictly equivalent to the traditional (Q)RPA in the small-amplitude limit (see Refs.~\cite{Scamps13a,Scamps14a} for a systematic study of both spherical and deformed systems). The multi-reference extension of time-dependent theories represents an alternative way to go beyond canonical (Q)RPA; see Ref.~\cite{Marevic23a} for a recent development in this direction.

A relevant effort relates to extending  \textit{ab initio} many-body methods, initially developed to address ground- and low-lying states~\cite{Hergert:2020bxy}, to describe GRs. For that purpose, hyperspherical harmonics (HH) and the no-core shell model (NCSM)~\cite{Quaglioni:2007biq,Quaglioni:2007eg,Bacca:2014tla} were first employed in light nuclei before the coupled-cluster (CC) method could be used to address closed-shell light and medium-mass nuclei~\cite{Bacca:2013dma,Bacca:2014rta,Bacca:2017yuj}. Very recently, the symmetry-adapted no-core shell model (SA-NCSM) has been used to further extend GR studies to open-shell medium-mass nuclei~\cite{Baker:2020rbq,Burrows:2023ugy}.  

The projected generator coordinate method (PGCM)~\cite{Ring80a} is a popular and versatile many-body method to study low-lying collective rotational and vibrational excitations based on phenomenological EDFs~\cite{bender03b,Niksic:2011sg,Robledo:2018cdj}\footnote{The PGCM is also employed as a low-cost alternative to shell-model diagonalisation in valence-space calculations~\cite{Gao:2015dla,Jiao:2017opc,Bally:2019miu,Shimizu:2021ltl,Sanchez-Fernandez:2021nfg}.}. The PGCM is able to explicitly account for so-called strong (i.e. {\it static}) correlations via the mixing of symmetry-breaking Slater determinants or Bogoliubov vacua  and via the subsequent restoration of the broken symmetries. Consequently, the PGCM is ideally suited to tackle the impact of nuclear superfluidity, deformation and shape mixing/coexistence in the description of collective states within a symmetry-conserving framework. Surprisingly though, the PGCM has been very seldom used for the description of GRs over the years~\cite{Caurier73a,Abgrall75a,Giraud75a,Flocard75a,Krewald76a,Flocard76a,Arickx83a,Stoitsov94a}. Still, it was observed long ago that anharmonic effects may have a non-negligible impact on the determination of the nuclear incompressibility~\cite{Blaizot95a}, which makes PGCM a useful tool in this respect.

Recently, the PGCM was adapted to the context of \textit{ab initio} calculations aiming at approximating exact solutions of many-body Schr{\"o}dinger's equation in the low-energy sector of the $A$-body Hilbert space starting from realistic nuclear Hamiltonians rooted into quantum chromodynamics~\cite{Frosini22a,Frosini22c}. In this setting, the PGCM delivers collective states that act as versatile unperturbed states on top of which a systematic expansion is performed to add complementary (so-called {\it dynamical}) correlations. This rationale was explicitly realised in the formulation of the PGCM-based many-body perturbation theory (PGCM-PT)~\cite{Frosini22a}. The first PGCM-PT calculations demonstrated that the PGCM itself delivers an excellent first approximation to the low-lying collective spectroscopy thanks to the nearly perfect cancellation of dynamical correlations between ground and excited states~\cite{Frosini22c}. This motivated the use of the PGCM as a standalone method\footnote{When proceeding to a pre-processing of the Hamiltonian via a multi-reference in-medium similarity renormalisation group transformation~\cite{Yao:2019rck,Yao:2018qjv,Frosini22c}, dynamical correlations on top the PGCM cancel to a lesser extent~\cite{Frosini22c}. Still, many avenues can be envisioned to better achieve such a cancellation~\cite{Duguet:2022zup}.} for the \textit{ab initio} study of low-lying collective states, indeed reaching a good reproduction of both experiment and quasi-exact solutions~\cite{Frosini22b}.

Inspired by such recent developments and extrapolating the cancellation of dynamical correlations to more excited collective states making up GRs, \textit{ab initio} PGCM calculations are presently employed to access the GMR (and the GQR) of light- and medium-mass for the first time.  Given that mid-mass nuclei can be systematically addressed independently of their doubly closed-shell, singly open-shell or doubly open-shell character, the PGCM nicely complements the the use of SA-NCSM~\cite{Baker:2020rbq,Burrows:2023ugy}. Eventually, the mean-field-like scaling with system's size of the PGCM computational cost makes it an excellent candidate to extend such studies to yet heavier closed- and open-shell nuclei in the future.

The present work is divided into four consecutive articles, hereafter coined as Paper I, Paper II~\cite{Porro24b}, Paper III~\cite{Porro24c} and Paper IV~\cite{Porro24d}. Paper I is presently dedicated to quantifying several uncertainty sources in the  \textit{ab initio} computation of the isoscalar monopole strength function via the PGCM, employing $^{46}$Ti as a typical example. Based on this analysis, Paper II discusses cases of physical interest, focusing on the mechanism determining the GMR-GQR coupling in intrinsically deformed systems. Next, Paper III analyzes different ways to access moments of the strength functions within PGCM. Finally, Paper IV focuses on the impact of restoring good angular momentum symmetry on the monopole strength function within PGCM calculations.

Paper I is organised as follows. Section~\ref{sec:theo} briefly introduces formal aspects of the PGCM and the computation of strength functions. In Sec.~\ref{sec:numerics} numerical aspects of the calculations are discussed. Section~\ref{UQ} constitutes the main part of the paper dedicated to the quantification of uncertainties in the PGCM calculations of the monopole and quadrupole strength functions. Eventually,  Sec.~\ref{sec:concl} provides the conclusions of Paper I.

\section{Formalism}
\label{sec:theo}

\subsection{PGCM ansatz}

The PGCM wave-function ansatz~\cite{Hill53a,Griffin57a} denotes a general superposition of so-called generating functions $\ket{\Phi(q)}$
\begin{align}
		\ket{\Psi_\nu^{\sigma}}
		&\equiv  \sum_q f^{\sigma}_{\nu}(q) P^\sigma | \Phi (q) \rangle \, , \label{eq:PGCM_ansatz1}
\end{align}
where $q$ denotes a set of collective variables called \textit{generator coordinates} and where $\sigma\equiv(JM\Pi NZ)$ characterises the symmetry quantum numbers carried by PGCM states, i.e. the angular momentum $J$ and its projection $M$, the parity $\Pi=\pm1$ as well as neutron $N$ and proton $Z$ numbers. The ensemble $\{ | \Phi(q) \rangle,\,q\in[q_0,q_1] \rangle\}$ denotes a set of non-orthogonal\footnote{Some Bogoliubov states mixed in the PGCM ansatz may be either manifestly or accidentally orthogonal. This situation can be dealt with at the price of a generalisation of the present work, where all pairs of Bogoliubov states entering Eq.~\eqref{eq:PGCM_ansatz1} are considered to be non-orthogonal.} Bogoliubov states typically obtained in a first step by repeatedly solving Hartree-Fock-Bogoliubov (HFB) mean-field equations with a Lagrange term associated with a constraining operator\footnote{The generic operator $Q$ typically embodies several constraining operators such that the collective coordinate $q$ is typically multi dimensional.} $Q$ such that the solution satisfies
\begin{align}
\langle \Phi(q) | Q | \Phi(q) \rangle &= q \, . \label{constraint}
\end{align}

In this process the state $| \Phi(q) \rangle$ typically breaks one or several symmetries of the Hamiltonian. Because physical states must carry good symmetry quantum numbers one acts on $| \Phi(q) \rangle$ in Eq.~\eqref{eq:PGCM_ansatz1} with the projection operator $P^\sigma$ associated with the symmetry (sub)group $\mathcal{G}$ of the Hamiltonian. The projection selects the component of $\ket{\Phi(q)}$ carrying the symmetry quantum numbers $\sigma$. In the case of present interest, $| \Phi(q) \rangle$ may break rotational and global-gauge symmetries associated with the conservation of angular momentum and nucleon numbers, respectively\footnote{While breaking rotational symmetry, presently employed Bogoliubov states still display axial symmetry, i.e. they do not display triaxial deformation and carry $M=0$ as a good quantum number.}. Consequently, the projector $P^\sigma$ explicitly restores both those symmetries. It can be generically written as 
\begin{equation}
	P^\sigma=\int\text{d}\varphi\,g^\sigma(\varphi)R(\varphi)\,, \label{projector}
\end{equation} 
where $g^\sigma(\varphi)$ represents a specific irreducible representation (IRREP) of $\mathcal{G}$ while $R(\varphi)$ denotes the unitary symmetry transformation operator changing the orientation of a state by the angle $\varphi$. Employing Eq.~\eqref{projector}, the PGCM state can be rewritten as
\begin{align}\label{eq:PGCM_ansatz}
	\ket{\Psi_\nu^{\sigma}}&= \sum_q \,f_\nu^\sigma(q)\sum_\varphi\,g^\sigma(\varphi)\ket{\Phi(q,\varphi)}\,,
\end{align}
where the $\varphi$-rotated state 
\begin{equation}
    \ket{\Phi(q,\varphi)}\equiv R(\varphi)\ket{\Phi(q)}
\end{equation}
has been introduced\footnote{If $\ket{\Phi(q)}$ is a Bogoliubov product state, $\ket{\Phi(q,\varphi)}$ is also a Bogoliubov product state. Indeed, the rotation operator $R(\varphi)$ can be represented as $R(\varphi)=e^{i\varphi S}$, where $S$ is a one-body operator and is referred to as the \textit{generator of the group}. By virtue of Thouless' theorem~\cite{thouless60}, $\ket{\Phi(q,\varphi)}$ is itself a Bogoliubov vacuum.} and where the integral over the rotation angle has been conveniently discretised.

\subsection{Hill-Wheeler-Griffin equation}

The unknown coefficients $f_\nu^\sigma(q)$ are determined variationally based on Ritz' variational principle, i.e. by minimising the energy associated with $| \Psi_\nu^\sigma \rangle$
\begin{equation}
	\delta\frac{\braket{\Psi^\sigma_\nu|H|\Psi^\sigma_\nu}}{\braket{\Psi^\sigma_\nu|\Psi^\sigma_\nu}}=0.
\end{equation}
The variation with respect to the weights $f_\nu^{\sigma\,*}(q)$ leads to solving a generalised eigenvalue problem known as the \textit{Hill-Wheeler-Griffin} secular equation~\cite{Griffin57a} 
\begin{equation}\label{eq:HWG}
	\sum_q\,\Big[\mathcal{H}^\sigma(p;q)-E^\sigma_\nu\mathcal{N}^\sigma(p;q)\Big]f_\nu^\sigma(q)=0\,,
\end{equation}
delivering the sequence of states $\{| \Psi^{\sigma}_\nu \rangle, \sigma \in \text{IRREPs}, \nu = 0, \ldots, \nu_{\text{max}} \}$ labelled by the symmetry quantum number $\sigma$ and the principal quantum number $\nu$\footnote{In principle, the number of states $\nu_{\text{max}}+1$ in each IRREP is equal to the cardinal $n_q$ of the ensemble $\{ | \Phi(q) \rangle,\,q\in[q_0,q_1] \rangle\}$. However, handling the  non-orthogonality of the states entering the PGCM ansatz leads in practice to an effective reduction of the dimensionality such that $\nu_{\text{max}}\leq n_q$.}. The secular equation involves the so-called symmetry-restored Hamiltonian and norm kernels 
\begin{subequations}\label{eq:kernels}
	\begin{align}
		\mathcal{H}^\sigma(p;q)&\equiv\braket{\Phi(p)|HP^\sigma|\Phi(q)}\nonumber\\
		&=\sum_\varphi\,g^\sigma(\varphi)\braket{\Phi(p)|H|\Phi(q,\varphi)}\,,\\
		\mathcal{N}^\sigma(p;q)&\equiv\braket{\Phi(p)|P^\sigma|\Phi(q)}\nonumber\\
		&=\sum_\varphi\,g^\sigma(\varphi)\braket{\Phi(p)|\Phi(q,\varphi)}\,, \label{eq:norm_ker}
	\end{align} 
\end{subequations}
themselves expressed in terms of the {\it unprojected} kernels 
\begin{subequations}
	\label{kernels}
	\begin{align}
		\mathcal{H}(p,q;\varphi) &\equiv \langle \Phi(p)|O|\Phi(q,\varphi) \rangle \, , \label{kernelO} \\
		\mathcal{N}(p,q;\varphi)&\equiv \langle \Phi(p)|\Phi(q,\varphi) \rangle \,, \label{kernelN}
	\end{align}
\end{subequations}
involving two Bogoliubov states carrying different values of the collective variables, the second state being further symmetry rotated. Eventually, the key ingredients to be computed are the unprojected norm kernel $\mathcal{N}(p,q;\varphi)$ and the {\it connected} Hamiltonian kernel~\cite{Duguet:2015yle,Bally:2017nom}  defined as 
\begin{align}
	{h}(p,q;\varphi) &\equiv  \frac{\mathcal{H}(p,q;\varphi)}{\mathcal{N}(p,q;\varphi)} \, . \label{connectedkernel}
\end{align}
The connected operator kernel  (Eq.~\eqref{connectedkernel}) can be efficiently computed by applying the off-diagonal Wick theorem (ODWT) of Balian and Brezin~\cite{BaBr69}. The evaluation of the norm kernel\footnote{For a detailed discussion about the connection between these different approaches and the hypothesis under which they are valid, see Refs.~\cite{Neergard:2022umu,Neergard:2023nox}.} (Eq.~\eqref{kernelN}) relies on the Onishi formula~\cite{onishi66}, the Pfaffian formula by Robledo~\cite{Rob09} or the integral formula by Bally and Duguet~\cite{Bally:2017nom}. Traditionally, the derivation of these two categories of kernels relies on different formal schemes that do not share a common ground. One exception relies on the use of fermion coherent states based on Grassmann variables allowing one to express both the connected operator kernel~\cite{Mizusaki:2012aq,Mizusaki:2013dda} and the norm kernel~\cite{Rob09} in terms of Pfaffians. Another consistent derivation of both kernels based on a diagrammatic method was recently proposed in Ref.~\cite{Porro22}.

\subsection{Intrinsic collective wave functions}

The linear redundancies due to the non-orthogonality of the HFB states mixed into the PGCM state must be dealt with when solving HWG’s equation. This is also a necessary step to extract meaningful intrinsic collective wave function as a function of the generator coordinates for each PGCM state. The detailed way to handle this problem can be found in, e.g. Ref~\cite{Frosini22b}.

Denoting as $\mathbf{N}^\sigma$ the Hermitian positive-definite matrix associated with the overlap kernel defined in Eq.~\eqref{eq:norm_ker}, it can be diagonalized according to
\begin{equation}
	\mathbf{N}^\sigma = \mathbf{S}^{\sigma\dagger}	\breve{\mathbf{N}}^\sigma \mathbf{S}^{\sigma}\,,
\end{equation}
where $ \mathbf{S}^{\sigma}$ is a unitary matrix and where $\breve{\mathbf{N}}^\sigma$ is diagonal with strictly positive eigenvalues. Defining
\begin{equation}
	\mathbf{G}^\sigma \equiv \mathbf{S}^{\sigma\dagger}	(\breve{\mathbf{N}}^\sigma)^{-1/2} \mathbf{S}^{\sigma}
\end{equation}
and only keeping the rows of $ \mathbf{S}^{\sigma}$ corresponding to eigenvalues of $\breve{\mathbf{N}}^\sigma$ larger than $\epsilon_{\text{th}}$, HWG’s equation~\eqref{eq:HWG} is transformed into the associated orthonormal basis and becomes
\begin{equation}
\sum_q\breve{H}^\sigma_{pq}\breve{f}^\sigma_\nu(q)=E_\nu^\sigma\breve{f}^\sigma_\nu(p)\,,
\end{equation}
with
\begin{subequations}
	\begin{align}
		\breve{\mathbf{H}}^\sigma & \equiv \mathbf{G}^{\sigma\dagger}\mathbf{H}^\sigma\mathbf{G}^{\sigma}\,,\\
		\mathbf{f}^\sigma & \equiv \mathbf{G}^{\sigma} \breve{\mathbf{f}}^{\sigma}\,.
	\end{align}
\end{subequations}
The solutions $\{ \breve{f}_\nu^\sigma(q); q \in \text{set}\}$ play the role of orthonormal intrinsic collective wave functions as a function of q that can be interpreted as probability amplitudes.

\subsection{GCM reduction}

The restoration of the symmetries broken by the Bogoliubov states  $\{ | \Phi(q) \rangle,\,q\in[q_0,q_1] \rangle\}$ is a key feature of the PGCM. Contrarily, the original GCM was formulated without taking it into consideration~\cite{Hill53a,Griffin57a}. Because it is the goal of Paper IV to investigate the impact of the angular momentum restoration on the monopole strength function, the GCM is also presently considered and can be easily obtained from the above by neglecting the projection operator $P^\sigma$. As a result, the GCM ansatz reduces to
\begin{equation}
	\ket{\Psi_\nu}=\sum_q\,f_\nu(q)\ket{\Phi(q)}\,.\label{eq:GCM_ansatz}
\end{equation}
The subsequent GCM HWG equation and GCM kernels can be straightforwardly obtained accordingly.

\subsection{Strength function and moments}

Given the set of PGCM states $\{| \Psi^{\sigma}_\nu \rangle, \sigma \in \text{IRREPs}, \nu = 0, \ldots, \nu_{\text{max}} \}$, the ground-state strength function associated with the transition operator $F$ is given by
\begin{equation}
    S_{F}(\omega)\equiv\sum_{\nu \sigma}|\braket{\Psi^{\sigma}_\nu|F|\Psi^{\sigma_0}_0}|^2 \, \delta(E^{\sigma}_\nu-E^{\sigma_0}_0-\omega) \, , \label{strengthfunction}
\end{equation}
where $| \Psi^{\sigma_0}_0 \rangle$ denotes the ground state. Moments of that strength function take the form\footnote{Paper III focuses on the evaluation of such moments within the PGCM.}
\begin{align}
m_k(F) &\equiv \sum_{\nu\sigma}(E^{\sigma}_\nu-E^{\sigma_0}_0)^k|\braket{\Psi^{\sigma}_\nu|F|\Psi^{\sigma_0}_0}|^2 \, ,\label{eq:def_k}
\end{align}
and thus require the knowledge of all excited states of the system. 

With these moments at hand, global characteristics of the strength functions, e.g. the centroid and the width can be respectively computed according to
\begin{subequations}
\begin{align}
    \bar{E}_1(F)&\equiv\frac{m_1(F)}{m_0(F)}\,,\\
    \tilde{E}_3(F)&\equiv\sqrt{\frac{m_3(F)}{m_1(F)}}\,,\\
    \tilde{E}_1(F)&\equiv\sqrt{\frac{m_1(F)}{m_{-1}(F)}}\,,
\end{align}
\label{eq:weighted_av}
\end{subequations}
and
\begin{equation}
\label{eq:stddev}
    \sigma(F)\equiv\sqrt{\frac{m_2(F)}{m_0(F)}-\Big(\frac{m_1(F)}{m_0(F)}\Big)^2}\, .
\end{equation}
While $\bar{E}_1$ actually denotes the centroid of the spectrum, $\tilde{E}_3$ and $\tilde{E}_1$ constitute alternative definitions of a mean energy, the former (latter) being more sensitive to the high (low) energy part of the strength. These three quantities coincide if the strength is located into a single peak. In addition to the width $\sigma$  characterising the fragmentation of the strength function, the hierarchy between $\bar{E}_1$, $\tilde{E}_3$ and $\tilde{E}_1$ allows one to quantify its asymmetry.

\section{Numerical aspects}
\label{sec:numerics}

The doubly open-shell nucleus $^{46}$Ti is used in the present paper to characterise PGCM calculations of the isoscalar GMR within an \textit{ab initio} setting.

\subsection{Expansion bases}

In the present calculations, a $k$-body operator is represented by a mode-$2k$ tensor expressed in the basis of the $k$-body Hilbert space built out of the $k$-fold tensor product of the one-body spherical harmonic oscillator (HO) basis. 

The one-body HO basis is defined via its frequency $\hbar\omega$ and is made finite dimensional by including all states up to $e_{\!_{\;\text{max}}}\equiv\text{max}(2n+l)$, with $n$ the principal quantum number and $l$ the orbital angular momentum. While two-body operators are consistently represented in a two-body basis including all states up to $e_{\!_{\;\text{2max}}}=2e_{\!_{\;\text{max}}}$, the representation of three-body operators is further restricted, due to computational limitations, by employing three-body states only up to $e_{\!_{\;\text{3max}}}<3e_{\!_{\;\text{max}}}$. In this work the value $e_{\!_{\;\text{3max}}}=14$ is systematically adopted, except for $e_{\!_{\;\text{max}}}=4$ give that the consistent value $e_{\!_{\;\text{3max}}}=12$  is tractable.

For large enough values of $(e_{\!_{\;\text{max}}},e_{\!_{\;\text{3max}}})$, observables do not carry any dependence on the chosen $\hbar\omega$. In practice, tractable values typically generate a residual $\hbar\omega$ dependence. In this context, certain frequencies become more optimal in the sense that observables are closer to their converged values. In the present work, values of $e_{\!_{\;\text{max}}}$ ($\hbar\omega$) ranging from 4 to 12 (12 to 24~MeV) are employed.

\subsection{Nuclear Hamiltonian}
\label{hamiltonians}

Several two- plus three-nucleon (2N+3N) chiral effective field theory ($\chi$EFT) based Hamiltonians are used in the present work. 

The family of Hamiltonians built in Ref.~\cite{Hu20} at next-to-leading-order (NLO), next-to-next-to-leading-order (N$^2$LO) and next-to-next-to-next-to-leading-order (N$^3$LO) is systematically employed. These Hamiltonians combine the 2N interaction by Entem and Machleidt (EM)~\cite{EnMa03,Machleidt11a} with a 3N force that is consistent concerning the chiral order, non-local regulator and cut-off value. The 2N low-energy constants (LECs) were fitted to two-nucleon scattering data, while 3N LECs were fixed using the ground-state energies of $^{3}$H and $^{16}$O. 

The NNLO$_{\text{sat}}$ Hamiltonian~\cite{Ekstrom15a} is further considered. It was introduced with the objective to provide improved saturation properties and radii, such that all its LECs were simultaneously fitted to few-body systems as well as to selected ground-state energies and radii of Carbon and Oxygen isotopes.

Eventually, the EM 1.8/2.0 interaction from Ref.~\cite{Nogga04a,Hebeler11a} is also employed. This Hamiltonian is based on the similarity-renormalisation-group (SRG) transformation of EM 2N interactions augmented with leading 3N contributions adjusted to data in $A=$3,4 systems. This Hamiltonian yields a saturation point of nuclear matter close to the empirical value~\cite{Hebeler11a} along with accurate binding energies in mid-mass nuclei~\cite{Stroberg21a}.

\subsection{SRG evolution}

Chiral EFT Hamiltonians are typically softer than phenomenological potentials used in seminal \textit{ab initio} calculations of light systems. This relates to the implicit treatment of high-energy degrees of freedom in chiral Hamiltonians, which is associated to a lower resolution scale and the absence of explicit coupling to high nucleon momenta.

Unitary SRG transformations can be employed to  further decouple low- and intermediate-momentum modes in the nuclear Hamiltonian~\cite{Jurgenson11a}. While mean-field HF or HFB solutions based on unevolved Hamiltonians are only shallowly bound (if not at all), the SRG evolution makes them more bound and consistently reduces many-body correlations to be added on top. In practice, the procedure leads to an acceleration of the convergence of the series at play in expansion many-body methods or may even turn a non-converging series into a converging one~\cite{Ti16}. At the same time, the decoupling of low- and intermediate-momentum modes authorises the use of smaller bases (i.e. $e_{\!_{\;\text{max}}}$ and $e_{\!_{\;\text{3max}}}$ values). The drawback of the SRG transformation resides in the appearance of additional three-body (and higher) contributions. In practice, beyond three-body interactions must be discarded, thus inducing a breaking of unitarity that needs to be controlled. 

The set of Hamiltonians from Ref.~\cite{Hu20} are presently evolved to two values of the low-momentum scale characterising the SRG transformation, namely $\lambda=1.88$~fm$^{-1}$ (i.e. flow parameter $\alpha$=0.08~fm$^4$) and $\lambda=2.24$~fm$^{-1}$ ($\alpha$=0.04~fm$^4$). 

The EM 1.8/2.0 Hamiltonian employs from the outset a 2N interaction evolved to $\lambda=1.8$~fm$^{-1}$ ($\alpha=0.0953$~fm$^4$), the regulator of the added 3N interaction being directly set to $\lambda=2.0$~fm$^{-1}$ without further explicit SRG evolution. As such, it belongs to the category of ``soft" nuclear Hamiltonians.

The NNLO$_{\text{sat}}$ Hamiltonian is presently kept unevolved and acts as a representative of ``hard" Hamiltonians displaying larger low-to-intermediate momentum couplings than the other ones.

\subsection{Three-body treatment}

Three-body forces (native or induced) are approximated via the rank-reduction method developed in Ref.~\cite{Frosini21a}. In practice the original or SRG-evolved Hamiltonian is first used to perform a spherical HFB calculation with the full treatment of two- and three-body operators. Three-body operators are then convoluted with the symmetry-conserving normal one-body density matrix thus computed in order to deliver a symmetry-conserving two-body effective interaction. This procedure reduces in closed-shell systems to the standard normal-ordered two-body approximation~\cite{Hebeler23a}.

\subsection{Generator coordinates}

The main objective is to compute the monopole strength function $S_{00}(\omega)\equiv S_{r^2}(\omega)$ obtained from Eq.~\eqref{strengthfunction} for $F\equiv r^2$, with the monopole operator defined as
\begin{align}
r^2 &\equiv \sum_{i=1}^\text{A} r_i ^2  \label{msrop} \, . 
\end{align}

The monopole response being mostly associated to radial vibrations of the nuclear surface, the root mean-square radius represents a natural choice for the main generator coordinate used in the calculation. Since it is known that intrinsic deformation may significantly affect $S_{00}(\omega)$ (see, for instance, Refs.~\cite{Peru08a,Gambacurta20a}), the  axial quadrupole deformation is the second generator coordinate of choice in the present study. To elaborate on such an impact, $S_{00}(\omega)$ is in fact analyzed along with the $K=0$ component $S_{20}(\omega)\equiv S_{Q_{20}}(\omega)$ of the quadrupole response\footnote{The evaluation of $K\neq0$ components requires the breaking of axial symmetry in the generation of the HFB vacua employed in the PGCM ansatz. The inclusion of the triaxial degree of freedom is not addressed in the present work and is left for future developments.}, the axial mass quadrupole operator being defined as
\begin{align}
Q_{20} &\equiv \sum_{i=1}^A r_i^2 \, Y_{20}(\vartheta_i, \phi_i) \, , \label{quadop}
\end{align}
where $(\vartheta_i, \phi_i)$ denote spherical angular coordinates. 

Eventually, most of the results presented below originate from two-dimensional (2D) PGCM calculations  $q~\equiv~(r,\beta_2)$ based on the set of Bogoliubov states  $\{ | \Phi(r,\beta_2) \rangle,\,r\in[r_{\text{min}},r_{\text{max}}], \beta_2 \in [\beta_{\text{min}},\beta_{\text{max}}] \rangle\}$ associated with the constraints\footnote{Note that the same notation $r^2$ is presently used to represent three different (although related) operators. While it is conventional to use the monopole operator introduced in Eq.~\eqref{msrop} to compute $S_{00}(\omega)$, the operator used in constrained HFB calculations (Eq.~\eqref{rms}) includes the prefactor $1/A$ given that it represents (a first approximation to) the nuclear radius. Whenever actually attempting to accurately compute the root-mean-square radius of a given many-body eigenstate, the center-of-mass correction is further added such that the employed operator becomes in this case
\begin{align}
r^2 &\equiv\frac{1}{\text{A}}\bigg(1-\frac{1}{\text{A}}\bigg)\sum_{i=1}^\text{A}r_i^2-\frac{2}{\text{A}^2}\sum_{i<j=1}^\text{A}\vec{r}_i\cdot\vec{r}_j\, . \label{2b}
\end{align}
}
\begin{subequations}
\begin{align}
r &\equiv \sqrt{\langle \Phi(r,\beta_2) | r^2 | \Phi(r,\beta_2) \rangle} \, , \label{rms} \\
\beta_2 &\equiv\frac{4\pi}{3R^2A}\langle \Phi(r,\beta_2) | Q_{20} | \Phi(r,\beta_2) \rangle \, ,  \label{beta2}
\end{align}
\end{subequations}
with $A$ the nucleon number, $R\equiv 1.2\,A^{1/3}$. On occasions, results from one-dimensional (1D) PGCM calculations based on $r$ as the sole coordinate will also presented.

\begin{figure*}[t]
    \centering
    \includegraphics[width=0.9\textwidth]{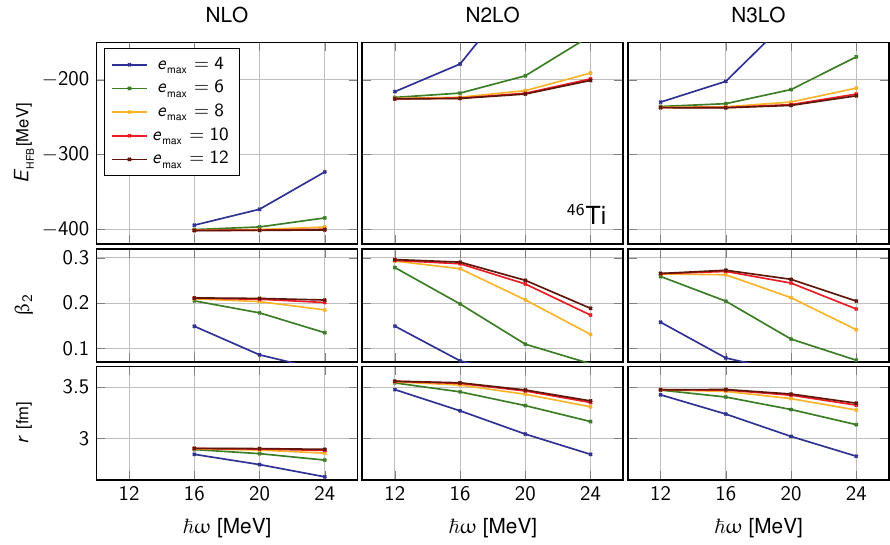}
    \caption{Total energy $E_{\text{HFB}}$ (Top row), axial deformation $\beta_2$ (middle row) and root-mean-square radius $r$ (bottom row) of the HFB minimum in $^{46}$Ti as a function of $\hbar\omega$ for several values of $e_{\!_{\;\text{max}}}$. Results employ NLO (left column), N$^2$LO (middle column) and N$^3$LO (right column) Hamiltonians from Ref.~\cite{Hu20} evolved to $\lambda_{\text{srg}}=1.88$~fm$^{-1}$.}
    \label{fig:HFB_chiral_Nav}
\end{figure*}

\subsection{Lorentzian smearing}

The resolution of the HWG equation (Eq.~\eqref{eq:HWG}) delivers the set of states $\{| \Psi^{\sigma}_\nu \rangle; \sigma \in \text{IRREPs}, \nu = 0, \ldots, \nu_{\text{max}}\}$ characterised by discrete eigenenergies. In order to increase the readability of multipole responses, discrete spectra are convoluted with a Lorentzian function of width $\Gamma$=0.5~MeV, unless specified otherwise, such that the strength function (Eq.~\eqref{strengthfunction}) is eventually replaced with its finite-resolution counterpart
\begin{equation}
    S_{F}(\omega)\equiv\sum_{\nu \sigma}|\braket{\Psi^{\sigma}_\nu|F|\Psi^{\sigma_0}_0}|^2 \, \frac{\Gamma}{(E^{\sigma}_\nu-E^{\sigma_0}_0-\omega)^2 + \Gamma^2}  \, .\label{strengthfunctionfinite}
\end{equation}
Given that experimental responses come, in most cases, with a finite energy resolution determined by an energy-bin width of about 500~keV, it is reasonable to take the finite resolution of the theoretical strength functions to be of the same order to proceed to their comparison.

\subsection{HFB minimum}
\label{sec:param_int}

To set up the stage for the following discussion, results obtained in $^{46}$Ti at the HFB minimum  using the family of chiral Hamiltonians from Ref.~\cite{Hu20} are presently characterised. The total energy $E_{\text{HFB}}$, its mean square radius $r$ and the axial deformation parameter $\beta_2$  are reported in Fig.~\ref{fig:HFB_chiral_Nav} as a function of $\hbar\omega$ for the three chiral orders and several $e_{\!_{\;\text{max}}}$ values. 

The energy displays a typical pattern as a function of the chiral order, i.e. the HFB binding energy decreases significantly going from NLO to N$^{2}$LO before increasing slightly when reaching N$^{3}$LO. 

Increasing $e_{\!_{\;\text{max}}}$, the convergence is observed at NLO for $e_{\!_{\;\text{max}}}=10$ for all quantities under consideration independently of  $\hbar\omega$. Contrarily, the convergence is reached at N$^{2}$LO and N$^{3}$LO for $e_{\!_{\;\text{max}}}=10$ only for low $\hbar\omega$ values, i.e. 12 and 16 MeV. This  different pattern is conjectured to be due to three-body forces that are indeed absent at NLO.

Eventually, Fig. \ref{fig:HFB_chiral_Nav} shows how HFB properties, in particular $r$ and $\beta_2$, can be severely affected by a non-optimal choice of $\hbar\omega$. As discussed in the following, this can impact significantly the determination of giant resonances. In general the $e_{\!_{\;\text{max}}}$ convergence is best achieved for $\hbar\omega=12$ MeV, which is an optimal choice for nuclei from $A\sim20$ to $A\sim 50$. This parameter is thus employed is all calculations below, unless stated otherwise.

\section{Uncertainty quantification}
\label{UQ}

The objective of Paper I is to provide a (partial) uncertainty quantification of ab initio PGCM calculations of the monopole response in mid-mass nuclei.  The sources of uncertainties are of three types
\begin{enumerate}
\item Hamiltonian modelling
\begin{enumerate}
\item Chiral EFT truncation order
\item LEC determination
\item Truncated SRG transformation
\end{enumerate}
\item Bases representation
\begin{enumerate}
\item One-body basis truncation $(e_{\!_{\;\text{max}}},\hbar \omega)$
\item Three-body basis truncation $(e_{\!_{\;\text{3max}}})$
\end{enumerate}
\item Many-body solution
\begin{enumerate}
\item Three-body interaction rank reduction
\item Choice of generator coordinates
\item Sampled values of generator coordinates 
\item Truncated PGCM-PT expansion 
\end{enumerate}
\end{enumerate}

In the present work
\begin{enumerate}
\item 1(b) and 3(b) are not evaluated,
\item 2(b), 3(a) and 3(d) are estimated based on previous works,
\item 1(c) is only touched upon,
\item 1(a), 2(a) and 3(c) are thoroughly evaluated.
\end{enumerate}
Ideally, all uncertainty sources must be systematically and consistently evaluated, especially given that several of them are all but independent. However, doing so is a daunting task and can only be the result of a long-term effort that goes well beyond the scope of the present work. For example, the propagation of the statistical uncertainty associated with the fit of the LECs in the employed $\chi$EFT Hamiltonian (1(b)) can be achieved on the basis of the eigenvector continuation (EC) techniques~\cite{Ekstrom:2019lss,Duguet:2023wuh}. The adaptation of the EC method to the PGCM is indeed currently underway~\cite{roux24a}.

\begin{figure*}
    \centering
    \includegraphics[width=0.9\textwidth]{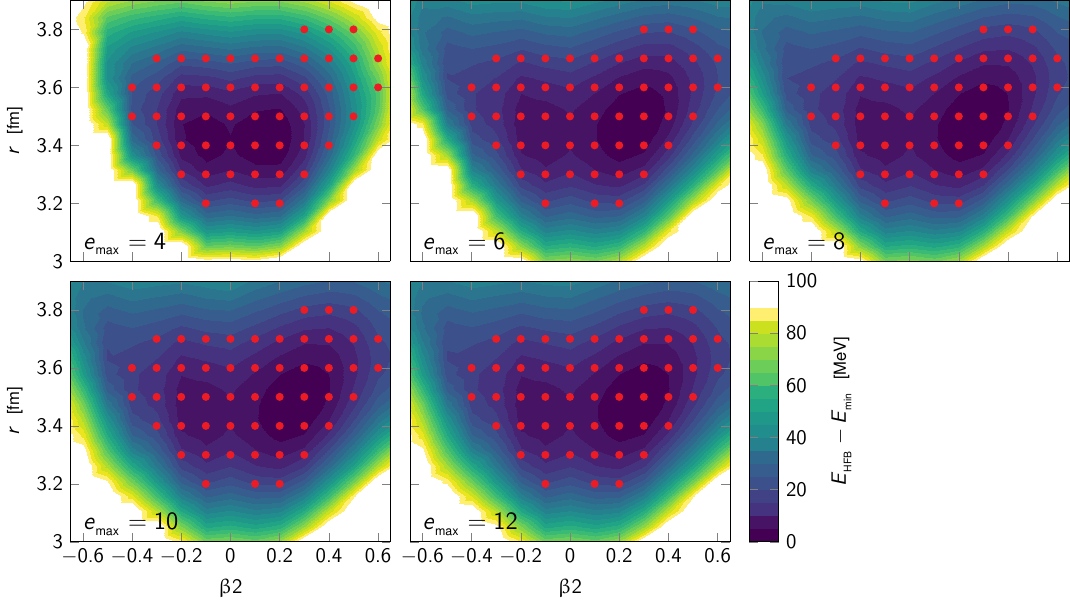}
    \caption{HFB total energy surface in $^{46}$Ti for different values of $e_{\!_{\;\text{max}}}$ ($\hbar\omega=12$ MeV) employing the N$^{3}$LO ($\lambda_{\text{srg}}=1.88$~fm$^{-1}$) Hamiltonian. The energy is plotted with respect to the minimal HFB energy for the corresponding $e_{\!_{\;\text{max}}}$ value.}
    \label{fig:PES_emax}
\end{figure*}

\begin{figure}[t]
    \centering
    \includegraphics[width=0.45\textwidth]{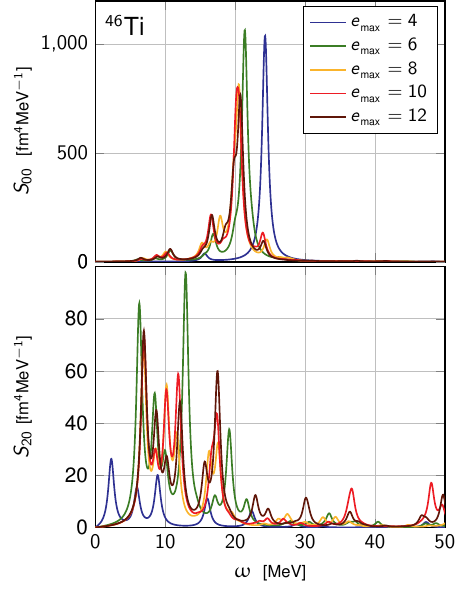}
    \caption{Monopole (top) and quadrupole (bottom) response in $^{46}$Ti employing the N$^{3}$LO ($\lambda_{\text{srg}}=1.88$~fm$^{-1}$) Hamiltonian for different values of $e_{\!_{\;\text{max}}}$ ($\hbar\omega=12$ MeV).}
    \label{fig:chiral_monopole}
\end{figure}

\subsection{Truncated PGCM-PT expansion}
\label{PGCMPTtrunc}

In the context of \textit{ab initio} calculations, the PGCM delivers the zeroth-order approximation to the recently formulated PGCM-PT expansion method~\cite{Frosini22a}. Contrary to standard, e.g. single-reference, expansion methods, such a zeroth-order state already captures a large fraction of the many-body correlations at play. Specifically, PGCM states incorporate strong static correlations associated to collective long-range fluctuations that are typically {\it hard} to grasp via standard expansion methods. The remaining dynamical, i.e. weak, correlations that come on top were shown to cancel essentially exactly in the excitation energy of low-lying rotational excitations when using low resolution-scale Hamiltonians~\cite{Frosini22c}. This motivated the use of the PGCM as a standalone method for the \textit{ab initio} study of low-lying collective excitations, indeed delivering excitation energies of low-lying collective natural-parity states consistent with quasi-exact solutions in several Ne isotopes~\cite{Frosini22b}. 

In this context, the present work relies on the assumption that such a feature remains valid for vibrational excitations making up the GMR in mid-mass nuclei, at least for its gross features and when employing low resolution-scale Hamiltonians. Confirming quantitatively to which extent this is indeed the case goes beyond the scope of the present work as it requires the extension of PGCM-PT to non-yrast states~\cite{bofos24a}.

\subsection{Three-body treatment}

The rank-reduction procedure presently applied to three-body interactions (native or induced)~\cite{Frosini21a} was shown to induce errors below 2-3\% across a large selection of nuclei, low-energy observables and many-body methods when based on low resolution-scale chiral Hamiltonians~\cite{Frosini21a}.

The actual evaluation of the error requires PGCM calculations of vibrational excitations making up the GMR in mid-mass nuclei with full 3N interactions, which is left for a future study. The present work relies on the assumption that the error remains on a few percent level as for the low-energy observables tested in Ref.~\cite{Frosini21a}.

\subsection{Three-body basis truncation}

For light- and mid-mass nuclei under present consideration, using $e_{\!_{\;\text{3max}}}=14$ is known to lead to a subpercent error for ground-state observables~\cite{Miyagi2021,Tichai:2023epe}. This conclusion remains to be tested for (highly) excited states such as those making up the GMR. This challenging point is however beyond the scope of the present study and is left for a future development.

\subsection{One-body basis truncation}
\label{sec:mod_space_conv}

In the present section, the  N$^{3}$LO Hamiltonian~\cite{Hu20}  evolved to $\lambda_{\text{srg}}=1.88$~fm$^{-1}$ is employed.
\begin{figure}
    \centering
    \includegraphics[width=0.4\textwidth]{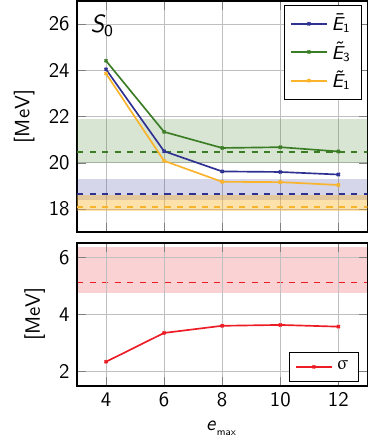}
    \caption{Mean energies (Eqs.~\eqref{eq:weighted_av}) and dispersion (\eqref{eq:stddev}) of the monopole response in $^{46}$Ti as a function of $\;e_{\!_{\;\text{max}}}$ ($\hbar\omega=12$ MeV) with the N$^{3}$LO ($\lambda_{\text{srg}}=1.88$~fm$^{-1}$) Hamiltonian. Dashed lines represent  experimental values, the shaded areas standing for the corresponding uncertainty~\cite{Tokimoto06a}.}
    \label{fig:mom_BE0_emax}
\end{figure}

\subsubsection{Dependence on $e_{\!_{\;\text{max}}}$}
\label{sec:emax}

As a first step, Fig.~\ref{fig:PES_emax} displays the 2D HFB total energy surface (TES) $E_{\text{HFB}}(r,\beta_2)$, rescaled to the HFB minimum, for different values of $e_{\!_{\;\text{max}}}$. Increasing the basis size decreases the stiffness of the TES when moving away from the minimum. By the time $e_{\!_{\;\text{max}}}=8$ is reached, the topology of the TES is stabilised. For any given point on the surface, the difference does not exceed $1$~MeV when going from $e_{\!_{\;\text{max}}}=10$ to $e_{\!_{\;\text{max}}}=12$. 

The red dots appearing in Fig.~\ref{fig:PES_emax} represent the HFB states included in the subsequent PGCM calculation. The corresponding monopole response $S_{00}(\omega)$ is displayed in the upper panel of Fig.~\ref{fig:chiral_monopole} for the different $\;e_{\!_{\;\text{max}}}$ values. The decreasing stiffness of the TES with $e_{\!_{\;\text{max}}}$ directly translates into the lowering of the GMR whose main peak eventually converges near $20.5$~MeV for $\;e_{\!_{\;\text{max}}}=8$. The finer structures visible over a wider energy range require $e_{\!_{\;\text{max}}}=10$ or even $e_{\!_{\;\text{max}}}=12$ to be stable. 

The quadrupole response $S_{20}(\omega)$ shown in the bottom panel of Fig.~\ref{fig:chiral_monopole} displays a slower convergence pattern than $S_{00}(\omega)$. While the overall structure is similar for $e_{\!_{\;\text{max}}}=10$ and $e_{\!_{\;\text{max}}}=12$, high-energy states are not converged. Similar trends in the quadrupole response are also observed in the next sections.

\begin{figure*}[t]
    \centering
    \includegraphics[width=0.95\textwidth]{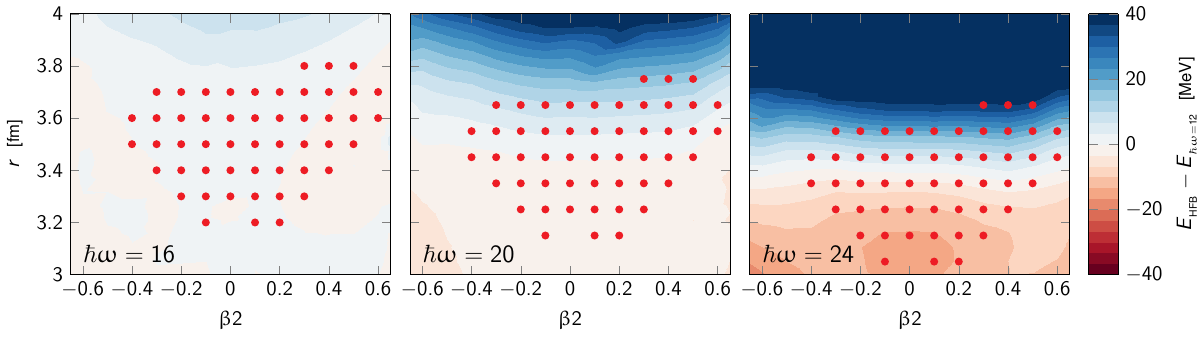}
    \caption{Same as Fig.~\ref{fig:PES_emax} for different values of $\hbar\omega$  ($e_{\!_{\;\text{max}}}=10$).  Energies are displayed with respect to the $\hbar\omega=12$ MeV results.}
    \label{fig:PES_hw_comp}
\end{figure*}
The convergence is now quantified by resorting to the main characteristics of the strength functions, i.e. the centroid and the dispersion obtained according to Eqs.~\eqref{eq:weighted_av} and~\eqref{eq:stddev}, respectively. Results are displayed in Fig.~\ref{fig:mom_BE0_emax} as a function of $e_{\!_{\;\text{max}}}$ for the monopole response. Experimental data from Ref.~\cite{Tokimoto06a} are shown as dashed lines, the shadowing representing the associated uncertainty. A clear convergence pattern is observed with $e_{\!_{\;\text{max}}}$. Based on this data set, the remaining uncertainty is estimated to be less than $1\%$ ($2\%$) for the centroid energy (dispersion).  

Increasing $e_{\!_{\;\text{max}}}$ significantly improves the agreement with experimental data. While the converged PGCM centroid value is $\sim$1~MeV higher than experiment, the dispersion underestimates it by about $1.4$~MeV. Including dynamical correlations beyond the strict PGCM along with the explicit coupling to the particle continuum is expected to improve the fragmentation of the strength. 

One may eventually note that, while $\tilde{E}_1$ is predicted to be close to the centroid value $\bar{E}_1$, $\tilde{E}_3$ is $1$~MeV higher, which results from the presence of several (small) peaks located well beyond the GMR. While this picture is consistent with experiment, the predicted difference $\tilde{E}_3-\bar{E}_1$ is only half of its experimental counterpart.

\subsubsection{Dependence on $\hbar \omega$}

The uncertainty associated with the choice of $\hbar\omega$ is now investigated at $\;e_{\!_{\;\text{max}}}=10$. As visible from Fig.~\ref{fig:PES_hw_comp}, increasing $\hbar\omega$ significantly modifies the topology of the TES with respect to $r$ compared to the optimal $\hbar\omega=12$~MeV value. As a result, and contrary to the previous study, the set of HFB states entering the PGCM ansatz evolves with $\hbar\omega$, i.e. it is shifted to lower values of $r$ with increasing $\hbar\omega$. 

\begin{figure}
    \centering
    \includegraphics[width=0.45\textwidth]{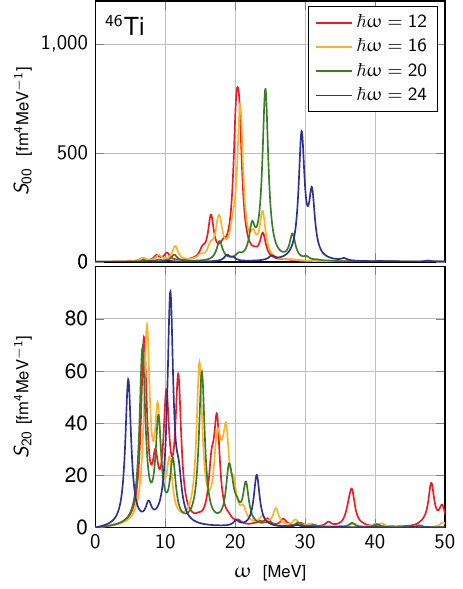}
    \caption{Same as Fig.~\ref{fig:chiral_monopole} for different values of $\hbar\omega$ ($\;e_{\!_{\;\text{max}}}=10$).}
    \label{fig:hw_monopole}
\end{figure}
\begin{figure}
    \centering
    \includegraphics[width=0.4\textwidth]{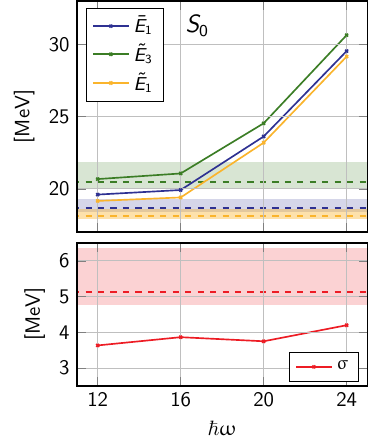}
    \caption{Same as Fig.~\ref{fig:mom_BE0_emax} for different values of $\hbar\omega$ ($\;e_{\!_{\;\text{max}}}=10$). Dashed lines represent experimental values, the shaded areas standing for the corresponding uncertainty~\cite{Tokimoto06a}.}
    \label{fig:mom_BE0_hw}
\end{figure}

Results for the monopole response are presented in the upper panel of Fig. \ref{fig:hw_monopole}. Going from $\hbar\omega=12$ to $\hbar\omega=16$ the resonance energy and the overall structure are barely modified. This reflects the mild differences between the two TES. At $\hbar\omega=20, 24$~MeV, the stiffer character of the TES moves the GMR to higher energies. 

The convergence pattern is less straightforward for the quadrupole response, even if lower-energy peaks are stable for $\hbar\omega=12-20$~MeV. Higher-energy structures appear for the optimal $\hbar\omega=12$~MeV case but not for the other $\hbar\omega$ values. 

As visible from Fig.~\ref{fig:mom_BE0_hw}, the centroid displays a gentle monotonic convergence going from $\hbar\omega=24$~MeV down to $\hbar\omega=12$~MeV, eventually sitting $1$~MeV above the experimental value as already discussed. The residual uncertainty is estimated to be about $1.5\%$. The dispersion rather slightly oscillates with $\hbar\omega$, systematically underestimating the experimental value. The residual uncertainty is estimated to be near $6\%$.

\begin{figure*}
    \centering
    \includegraphics[width=0.95\textwidth]{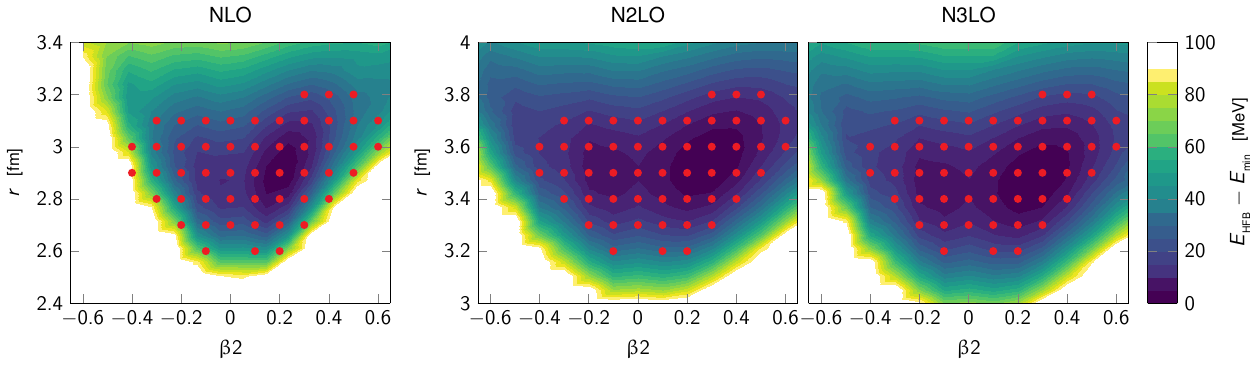}
    \caption{Same as Fig.~\ref{fig:PES_emax} for $(e_{\!_{\;\text{max}}}, \hbar\omega)=(10, 16\,\text{MeV})$ employing Hamiltonians at different chiral orders ($\lambda_{\text{srg}}=1.88$~fm$^{-1}$). The energy is plotted with respect to the minimal HFB energy for the respective chiral order.}
    \label{fig:PES_chiral}
\end{figure*}

\subsection{Chiral EFT truncation order}
\label{sec:conv_chiral}

The convergence pattern with respect to the chiral order of the employed Hamiltonian was displayed in Fig.~\ref{fig:HFB_chiral_Nav} for the total HFB energy, radius and $\beta_2$ parameter.
These quantities were shown to be strongly modified when going from NLO to N$^{3}$LO, while they undergo a milder change going from N$^{2}$LO to N$^{3}$LO. 

\begin{figure}
    \centering
    \includegraphics[width=0.45\textwidth]{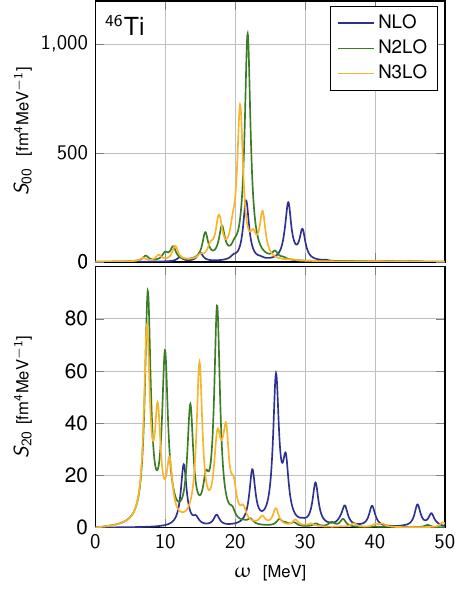}
    \caption{Same as Fig.~\ref{fig:hw_monopole} for $(e_{\!_{\;\text{max}}}, \hbar\omega)=(10, 16\,\text{MeV})$ employing Hamiltonians at different chiral orders ($\lambda_{\text{srg}}=1.88$~fm$^{-1}$).}
    \label{fig:spectra_chiral}
\end{figure}
The HFB TES is shown in Fig.~\ref{fig:PES_chiral} for the three chiral orders. Given the smaller HFB radius at NLO, the TES is shifted by approximately $-0.6$ fm with respect to N$^{2}$LO and N$^{3}$LO. The NLO TES is also significantly stiffer. 

\begin{figure}
    \centering
    \includegraphics[width=0.4\textwidth]{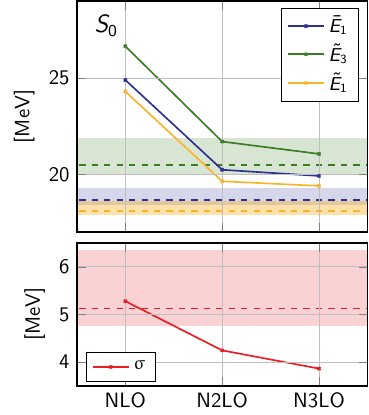}
    \caption{Same as Fig.~\ref{fig:mom_BE0_hw} for $(e_{\!_{\;\text{max}}}, \hbar\omega)=(10, 16\,\text{MeV})$ employing Hamiltonians at different chiral orders ($\lambda_{\text{srg}}=1.88$~fm$^{-1}$). Dashed lines represent the corresponding experimental values, the shaded areas standing for the corresponding uncertainty~\cite{Tokimoto06a}.}
    \label{fig:mom_BE0_chiral}
\end{figure}

Properties of the HFB TES directly translate into the monopole responses displayed in Fig.~\ref{fig:spectra_chiral} (top panel). NLO results are very distinct from N$^{2}$LO and N$^{3}$LO ones, the stiffer TES of the former pushing the GMR $8$~MeV higher in energy. Moreover, the magnitude of the strength is significantly smaller than at N$^{3}$LO. Going from N$^{2}$LO to N$^{3}$LO, the position of the GMR is further lowered by $\sim$1~MeV.

Similar observations concern the quadrupole response in Fig.~\ref{fig:spectra_chiral} (bottom panel). The NLO strengh function exhibits little resemblance to N$^{2}$LO and N$^{3}$LO responses. The latter two are similar even though the highly fragmented response below 20~MeV is still significantly modified going from N$^{2}$LO to N$^{3}$LO.


The mean energies extracted from the monopole response are shown in the upper panel of Fig.~\ref{fig:mom_BE0_chiral} to display a converging pattern with the chiral order, the N$^{3}$LO results comparing favorably with experimental data. Based on this pattern, the remaining uncertainty associated with omitted higher orders is estimated to be below $2\%$.

While the dispersion changes less going from N$^{2}$LO to N$^{3}$LO than from NLO to N$^{2}$LO, the converging pattern is not obvious. Assuming (naively) that the dispersion does converge monotonically, the residual uncertainty can be estimated to be on a $10\%$ level. This confirms that the PGCM value is inconsistent with experiment\footnote{The fact that the NLO dispersion agrees well with experiment is accidental and masks the fact that the NLO spectrum does not display any giant resonance.} due to missing many-body (dynamical) correlations\footnote{Since the two uncertainties are not independent, incorporating  dynamical correlations would not only change the central value of the dispersion but also the associated chiral truncation uncertainty.}  

\subsection{Truncated SRG transformation}
\label{sec:srg}

The evolution of the monopole response in PGCM calculations with respect to the SRG flow parameter is now addressed. Fully converged \textit{ab initio} observables should not carry any explicit dependence on the SRG evolution, which is in principle a unitary transformation of the initial Hamiltonian. In this sense, the residual dependence on the SRG scale is an indicator of the breaking of unitarity in practical applications. This breaking is a convolution of the truncation of the transformed Hamiltonian and of the approximation made on the computation of its many-body eigenstates. 

In this context, results obtained for two different values ($\alpha=0.04$ fm$^4$ and $\alpha=0.08$ fm$^4$) of the flow parameters are presently compared to give a {\it sense} of the breaking of unitarity and of the associated uncertainty.

\begin{table}
    \centering
    \begin{tabular}{crcl}
    \hline
        $\alpha$ [fm$^4$]  & $E_{\text{HFB}}$ [MeV] & $\beta_2$ & $r$ [fm] \\ \hline
        0.04 & -145.895 & 0.28 & 3.65 \\ 
        0.08 & -237.299 & 0.26 & 3.47 \\ \hline
    \end{tabular}
    \caption{Total energy $E_{\text{HFB}}$, axial deformation $\beta_2$ and root mean square radius $r$ of the HFB minimum in $^{46}$Ti ($\hbar\omega=12$~MeV and $e_{\!_{\;\text{max}}}=10$) for different values of the flow parameter $\alpha$.}
    \label{tab:HFB_srg}
\end{table}

\begin{figure}
    \centering
    \includegraphics[width=0.48\textwidth]{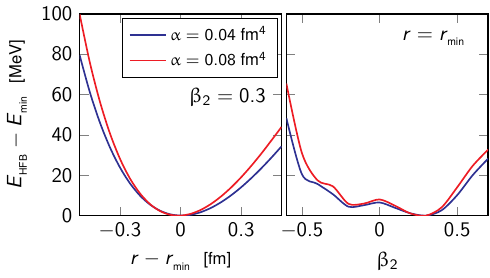}
    \caption{One-dimensional cuts of the total energy surface of $^{46}$Ti  for different values of the flow parameter ($\hbar\omega=12$ MeV and $e_{max}=10$). Left panel: energy as a function of $r$ for the $\beta_2$ value minimising the HFB energy. Right panel: energy as a function of $\beta_2$ for the $r$ value minimising the HFB energy. Calculations employ the N$^{3}$LO Hamiltonian and the energy is plotted with respect to the minimal HFB energy for the respective flow parameter.}
    \label{fig:PES_srg_1d_cut}
\end{figure}
The change of observables at the HFB minimum are reported in Tab.~\ref{tab:HFB_srg}. It is observed that softening the Hamiltonian from  $\alpha=0.04$ fm$^4$ to $\alpha=0.08$ fm$^4$ lowers the HFB energy by  about 100~MeV and reduces the root-mean-square radius by $0.17$~fm.

\begin{figure}
    \centering
    \includegraphics[width=0.45\textwidth]{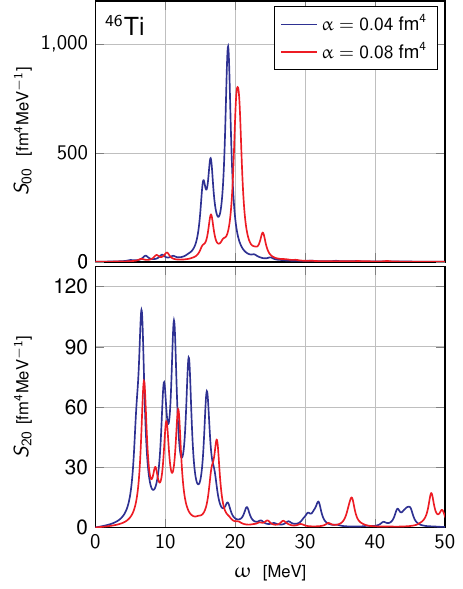}
    \caption{Same as Fig.~\ref{fig:hw_monopole} for different values of the flow parameter at N$^{3}$LO ($\hbar\omega=12$~MeV and $\;e_{\!_{\;\text{max}}}=10$).}
    \label{fig:spectra_srg}
\end{figure}

\begin{figure}
    \centering
    \includegraphics[width=0.4\textwidth]{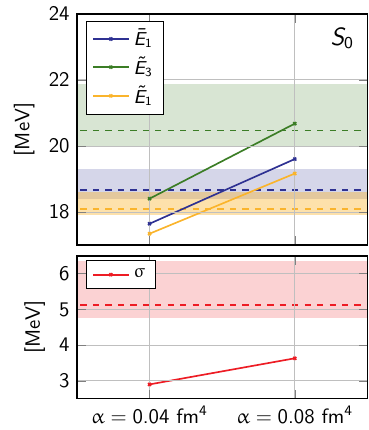}
    \caption{Same as Fig.~\ref{fig:mom_BE0_hw} for different values of the flow parameter at N$^{3}$LO ($\hbar\omega=12$~MeV and $\;e_{\!_{\;\text{max}}}=10$). Dashed lines represent the  experimental values,  the shaded areas standing for the corresponding uncertainty~\cite{Tokimoto06a}.}
    \label{fig:mom_srg}
\end{figure}

In addition to lowering the mean-field TES altogether by resumming dynamical correlations into the Hamiltonian, the SRG transformation (i.e. the increase of the flow parameter $\alpha$) particularly favors the ground-state, i.e. the HFB minimum. As a result, the TES becomes stiffer as is visible from Fig.~\ref{fig:PES_srg_1d_cut} displaying two 1D cuts of the 2D TES through the HFB minimum: the dependencies on $r$ and $\beta_2$ are both slightly stiffer for $\alpha=$0.08 fm$^4$. This observation straightforwardly translates into the behaviour of the monopole and quadrupole responses shown in Fig.~\ref{fig:spectra_srg}. The two monopole responses (top panel) display a qualitatively similar behaviour, the $\alpha=$0.08~fm$^4$ strength being shifted up by $\sim$1~MeV compared to the $\alpha=$0.04 fm$^4$ one. 

While the quadrupole response is more complex, the effect of the truncated SRG transformation is similar to the one observed on the monopole strength as the bottom panel of Fig.~\ref{fig:spectra_srg} testifies. Interestingly the high-energy structures between 30 and 50 MeV appear in both calculations and are also shifted up going from $\alpha=$0.04 fm$^4$ to $\alpha=$0.08 fm$^4$.



The anti-correlation between the softening of the Hamiltonian and the position of the monopole and quadrupole centroids was also observed in RPA calculations of spherical systems~\cite{Tr16}. Such a feature is further quantified in the upper panel of Fig.~\ref{fig:mom_srg} for the monopole response. One observes that the same is true for the dispersion. 

Eventually, the use of two $\alpha$ values is not sufficient to extract a quantitative uncertainty associated with the truncated SRG transformation. Furthermore, this error is strongly convoluted with the uncertainty associated with the truncated PGCM-PT expansion such that any realistic assessment should involved both aspects at the same time. One can only observe that relying on the strict PGCM is probably delivering a better estimate of the exact solution when combined with a soft Hamiltonian ($\alpha=$0.08 fm$^4$), as done by default in the present work. Indeed, in such a case dynamical correlations on top of the PGCM (i) are smaller in absolute as illustrated above with the lowering of the HFB energy and (ii) have been shown to perfectly cancel out in the excitation energy of low-lying collective states~\cite{Frosini22c}. While this does not constitute a proper uncertainty quantification, it gives credit to the results obtained in the present study.

\subsection{Comparison of different Hamiltonians}
\label{sec:compare_int}

In addition to the systematic uncertainty associated with the truncated chiral order and SRG transformations, the Hamiltonian carries a statistical uncertainty associated with the adjustment of its LECs that needs to be propagated to many-body observables. As mentioned above, the latter is however not evaluated in the present work. 

In this context, a poor-man's way to qualify the sensitivity of the results to the input Hamiltonian consists of employing a set of representative Hamiltonians differing in their Chiral order, SRG scale and/or fitting protocol. In this spirit, three different chiral-based nuclear Hamiltonians are tested in the present section, i.e. the N$^{3}$LO \cite{Hu20}, NNLO$_{\text{\,sat}}$ \cite{Ekstrom15a} and EM 1.8/2.0 \cite{Nogga04a,Hebeler11a} Hamiltonians whose characteristics were briefly detailed in Sec.~\ref{hamiltonians}.
\begin{table}
    \centering
    \begin{tabular}{rrcl}
    \hline
        ~ & $E_{\text{HFB}}$ [MeV] & $\beta_2$ & $r$ [fm] \\ \hline
        N$^{3}$LO & -237.30 & 0.27 & 3.47 \\ 
        NNLO$_{\text{\,sat}}$ & -95.46 & 0.34 & 3.72 \\ 
        EM 1.8/2.0 & -284.24 & 0.24 & 3.32 \\ \hline
    \end{tabular}
    \caption{Total energy $E_{\text{HFB}}$, axial deformation $\beta_2$ and root-mean-square radius $r$ of the HFB minimum in $^{46}$Ti for different chiral-based Hamiltonians  ($\hbar\omega=12$~MeV and $e_{\!_{\;\text{max}}}=10$).}
    \label{tab:HFB_int}
\end{table}

\begin{figure}
    \centering
    \includegraphics[width=0.48\textwidth]{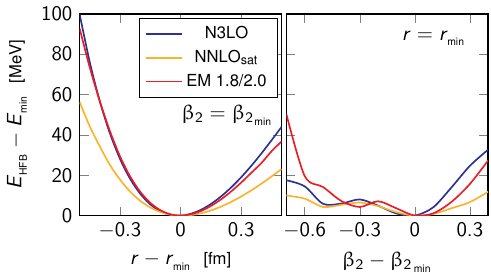}
    \caption{Same as Fig.~\ref{fig:PES_srg_1d_cut} for the N$^{3}$LO, NNLO$_{\text{\,sat}}$ and EM 1.8/2.0  Hamiltonians ($\hbar\omega=12$~MeV and $\;e_{\!_{\;\text{max}}}=10$).}
    \label{fig:PES_int_1d_cut}
\end{figure}

The optimal basis parameters ($e_{\!_{\;\text{max}}}=10$ and $\hbar\omega=12$) are found to be the same for the three Hamiltonians. Characteristics of the HFB minimum are listed in Tab.~\ref{tab:HFB_int}. The NNLO$_{\text{\,sat}}$ Hamiltonian being used without SRG evolution, the system is very shallowly bound at the HFB level. The N$^{3}$LO (EM 1.8/2.0) Hamiltonian being evolved (further) down to the resolution $\alpha=0.08$ fm$^4$ ($\alpha=$0.0953 fm$^4$), the HFB energy is lowered by 130~MeV (180~MeV). Radii predictions are poorer for EM 1.8/2.0, as already known from the literature~\cite{Lapoux16a}.

\begin{figure}
    \centering
    \includegraphics[width=0.45\textwidth]{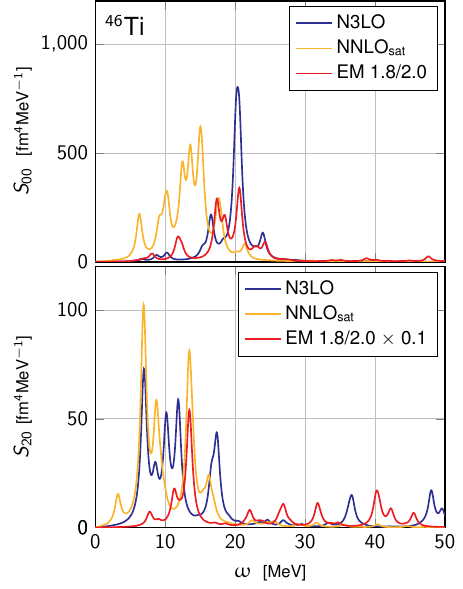}
    \caption{Same as Fig.~\ref{fig:hw_monopole} for the N$^{3}$LO, NNLO$_{\text{\,sat}}$ and EM 1.8/2.0  Hamiltonians ($\hbar\omega=12$~MeV and $\;e_{\!_{\;\text{max}}}=10$).}
    \label{fig:spectra_int}
\end{figure}

\begin{figure*}
    \centering
    \includegraphics[width=0.95\textwidth]{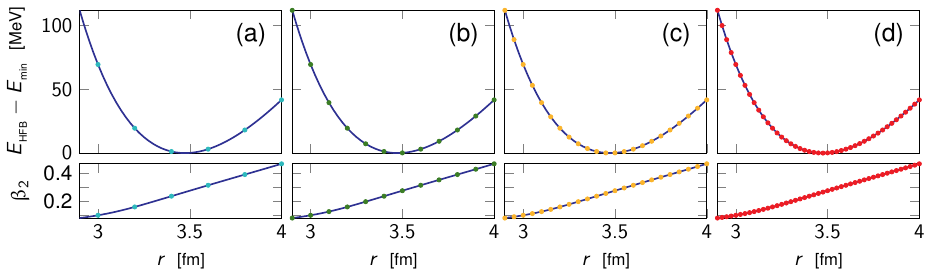}
    \caption{HFB one-dimensional total energy surface in $^{46}$Ti using $r$ as the generator coordinate. The N$^{3}$LO ($\alpha=$0.08 fm$^4$) Hamiltonian is employed ($\hbar\omega=12$~MeV and $\;e_{\!_{\;\text{max}}}=10$). The mesh discretization in $r$ for the subsequent PGCM calculation is (a) $0.2$ fm (6 points), (b) $0.1$ fm (12 points), (c) $0.05$ fm (23 points) and (d) $0.025$ fm (45 points).}
    \label{fig:1d_PES}
\end{figure*}

\begin{figure}
    \centering
    \includegraphics[width=0.45\textwidth]{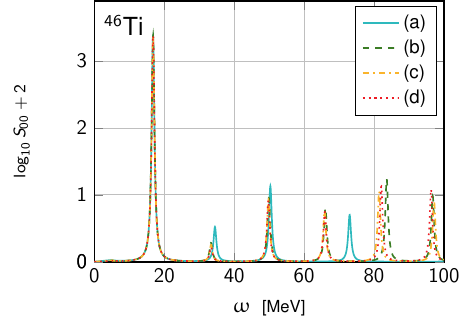}
    \caption{Monopole responses in $^{46}$Ti computed with the N$^{3}$LO ($\alpha=$0.08 fm$^4$) Hamiltonian  ($\hbar\omega=12$~MeV and $\;e_{\!_{\;\text{max}}}=10$). The sets of HFB states entering the PGCM ansatz correspond to the points appearing on the four one-dimensional total energy surfaces shown in Fig.~\ref{fig:1d_PES}. A logarithmic scale is employed for the vertical axis. See Fig. \ref{fig:1d_PES} for details on labels and calculation parameters.}
    \label{fig:1d_spectra}
\end{figure}

Compared to N$^{3}$LO, the NNLO$_{\text{\,sat}}$ TES favors large deformations and is extremely soft against $\beta_2$ and $r$. The EM 1.8/2.0 TES is similar to the N$^{3}$LO one, even though it is shifted to smaller radii by $\sim-0.2$ fm. A quantitative statement about the stiffness of the TES is inferred from Fig.~\ref{fig:PES_int_1d_cut}, where 1D cuts through the HFB minimum are shown. As far as radial variations are concerned, N$^{3}$LO and EM 1.8/2.0 behave very similarly. In contrast, NNLO$_{\text{\,sat}}$ is much softer with respect to both compression and dilatation. According to the conclusions of Secs.~\ref{sec:conv_chiral} and~\ref{sec:srg}, this characteristic reflects more the unevolved character of NNLO$_{\text{\,sat}}$ than the different chiral order.

These patterns leave their fingerprint on the monopole responses displayed in the top panel of Fig.~\ref{fig:spectra_int}. Taking N$^{3}$LO as a reference, NNLO$_{\text{\,sat}}$ produces a highly fragmented monopole response, overall shifted down by about 10~MeV. Because the rigidity of the EM 1.8/2.0 PES with respect to $r$ is similar to N$^{3}$LO, the monopole response remains localised within the same energy interval, i.e. between 12 and 25~MeV. It is however more fragmented possibly due to a different coupling to the quadrupole deformation.

The behaviour of the quadrupole response is different, as seen in the bottom panel of Fig.~\ref{fig:spectra_int}. Similarly to N$^{3}$LO, NNLO$_{\text{\,sat}}$ generates a fragmented response below 20~MeV, although shifted towards lower energies due to the softer TES against $\beta_2$ as visible in the right panel of Fig.~\ref{fig:PES_int_1d_cut}. Oddly, the quadrupole response of EM 1.8/2.0 is extremely large (notice the 0.1 factor) and can hardly be related to the other results or inferred from Fig.~\ref{fig:PES_int_1d_cut}, except for its location at higher energy that can be understood from its stiffer TES against $\beta_2$.

Eventually, results vary significantly with the employed Hamiltonian. While the use of unevolved interactions such as NNLO$_{\text{\,sat}}$ at the strict PGCM level is probably unsafe, it is hard to rationalise the difference between N$^{3}$LO and EM 1.8/2.0, especially the behavior of the latter in the quadrupole channel. In the future, robust model uncertainties combining consistent variations in the Hamiltonian modelling (e.g. chiral order, momentum regulators, cutoff values, LECs uncertainties, fitting protocol, SRG transformation) should be pursued to propagate uncertainties to, e.g., PGCM predictions.




\subsection{Choice of generator coordinates}
\label{sec:pgcm_choice_coordinates}

The unique strength of the PGCM relates to its capacity to treat collective fluctuations and incorporate associated strong static correlations. This is achieved via a physically-intuitive selection of relevant generator coordinates. However, such a strength turns into a shortcoming when it comes to evaluating systematic uncertainties. In practice, the generator coordinates are chosen by the practitioner based on the physics to be addressed. For example, $r$ and $\beta_2$ were presently chosen as the minimal set necessary to (i) describe the monopole resonance (ii) for systems potentially displaying an intrinsic quadrupole deformation. While such a choice is well motivated empirically, the procedure provides the PGCM with a lack of systematic character. 

Eventually, evaluating the uncertainty associated with the choice of generator coordinates is difficult given that realistic state-of-the-art calculations cannot be performed with more than a few collective coordinates in practice. Furthermore, within the frame of \textit{ab initio} calculations, evaluating this uncertainty cannot be done independently of the one associated with the truncation of the PGCM-PT expansion itself. Indeed, static and dynamical correlations are not orthogonal to one another such that a reshuffling operates between both categories depending on the nature of the PGCM unperturbed state, i.e. depending on the choice of collective coordinates\footnote{The statement made in Sec.~\ref{PGCMPTtrunc} about the cancellation of dynamical correlations in the excitation energy of low-lying collective states is in fact generator-coordinates-dependent and thus needs to be further explored in the future while varying the set of generator coordinates defining the PGCM ansatz.}.

\begin{figure*}
    \centering
    \includegraphics[width=0.95\textwidth]{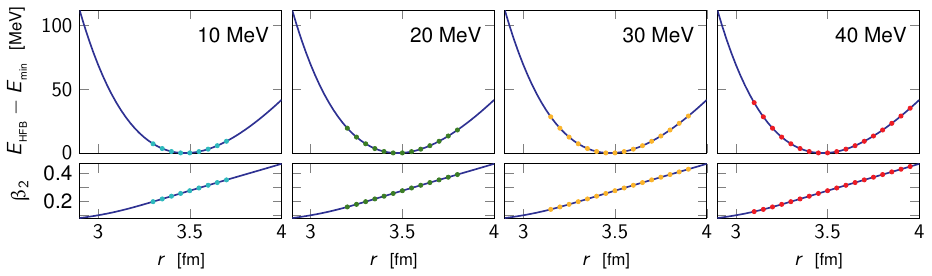}
    \caption{Same as Fig.~\ref{fig:1d_PES} except that the HFB states included in the subsequent PGCM calculation are selected according to four different values of $E_{\text{max}}$, 10 MeV (9 points), 20 MeV (13 points), 30 MeV (16 points) and 40 MeV (18 points) for the fixed equidistant mesh value of $0.05$~fm.}
    \label{fig:1d_PES_E}
\end{figure*}

\begin{figure}
    \centering
    \includegraphics[width=0.45\textwidth]{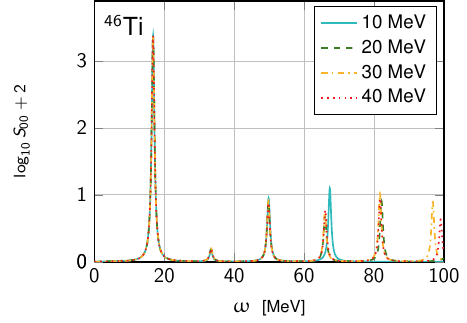}
    \caption{Same as Fig.~\ref{fig:1d_spectra} except that the HFB states entering the  PGCM ansatz are selected according to four increasing values of $E_{\text{max}}$, i.e. 10, 20, 30 and 40~MeV for a fixed equidistant mesh value of $0.05$~fm.}
    \label{fig:1d_spectra_E}
\end{figure}

For these reasons, much developments are needed in order to be in the position to evaluate this uncertainty in a systematic way\footnote{In fact, the very way the Bogoliubov states are generated as a function of the generator coordinate $q$ needs to be  investigated systematically. While it has been customary to select adiabatic vacua obtained by solving constrained HFB equations, other avenues can be envisioned, e.g. choosing non-adiabatic states carrying the constrained deformation $q$.} and is thus beyond the scope of the present work.

\subsection{Sampled values of generator coordinates}
\label{sec:pgcm_points_conv}

Given a generator coordinate $q$, one must still select a discrete set of values $q_i \in [q_1,q_2], i =1,\ldots,n_q$ to perform PGCM calculations. However, there is no unique criterion to perform such a selection efficiently (see~\cite{Martinez22a} for a recent study), i.e. in a way that suppresses the dependence of the results on this selection.

In this section, two criteria are used to select the HFB states entering the PGCM ansatz and evaluate the uncertainty associated with it. The first one relates the step size of the uniform mesh employed while the second one concerns the spanned interval $[q_1,q_2]$ that is translated into a maximum energy difference $E_{\text{max}}$ with respect to the HFB minimum, i.e. $E_{\text{HFB}}(r,\beta_2) - E_{\text{min}} \leq E_{\text{max}}$. The uncertainty associated with the two selection criteria is evaluated sequentially for 1D and 2D PGCM calculations. Calculations are performed using the N$^{3}$LO ($\alpha=$0.08 fm$^4$) Hamiltonian with $\hbar\omega=12$~MeV and $\;e_{\!_{\;\text{max}}}=10$. 

\subsubsection{One-dimensional case}
\label{sec:pgcm1d_calc}

A 1D PGCM is first employed to limit the complexity of the analysis. Calculations are  performed solely constraining the generator coordinate $r$.

\paragraph{Mesh resolution}

Four (identical) 1D HFB TES as a function of $r$ are shown in the upper panels of Fig.~\ref{fig:1d_PES}. They each display a different set of HFB states used in the subsequent PGCM calculation according to mesh step size (using a fixed interval), i.e. (a) $0.2$ fm, (b) $0.1$ fm, (c) $0.05$ fm and (d) $0.025$ fm. This corresponds to performing PGCM calculations based on 6, 12, 23 and 45 HFB states, respectively, i.e. doubling the mesh density at each step going from left to right. 

Monopole responses from the four 1D PGCM calculations are displayed in Fig.~\ref{fig:1d_spectra} using a logarithmic scale. Strikingly, results are independent of the chosen mesh such that no difference would be visible using a linear scale. Even  high-energy peaks carrying tiny transition amplitudes only differ very slightly using the four different step sizes. 

\begin{figure*}
    \centering
    \includegraphics[width=0.95\textwidth]{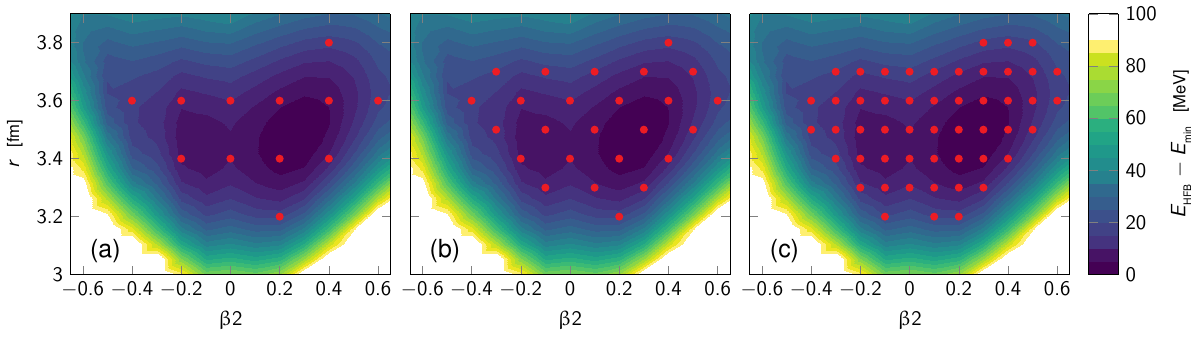}
    \caption{HFB total energy surfaces in $^{46}$Ti using the N$^{3}$LO ($\alpha=$0.08 fm$^4$) Hamiltonian ($\hbar\omega=12$ MeV and $\;e_{\!_{\;\text{max}}}=10$). Each case displays the set of HFB states included in the subsequent PGCM calculation according to a chosen mesh. The chosen meshes are denser going from left to right and correspond to using (a) 12 points, (b) 25 points and (c) 51 points, respectively.}
    \label{fig:PES_2d_points}
\end{figure*}

\begin{figure}
    \centering
    \includegraphics[width=0.45\textwidth]{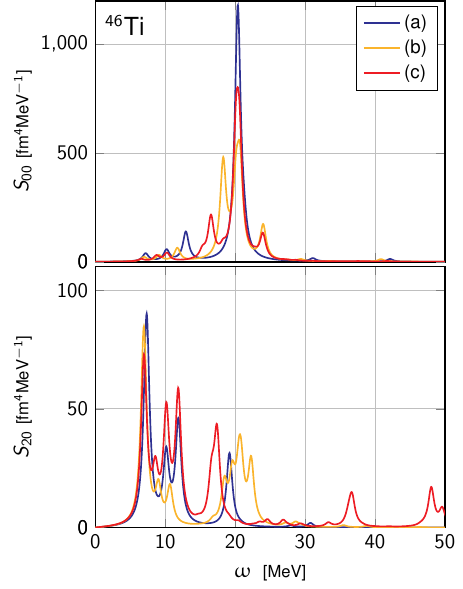}
    \caption{Monopole (top) and quadrupole (bottom) responses in $^{46}$Ti based on the N$^{3}$LO ($\lambda_{\text{srg}}=1.88$~fm$^{-1}$) Hamiltonian ($e_{\!_{\;\text{max}}}=10, \hbar\omega=12$~MeV)  and computed using the three sets of HFB states displayed in Fig.~\ref{fig:PES_2d_points} into the PGCM ansatz.}
    \label{fig:spectra_2d_points}
\end{figure}

\begin{figure}
    \centering
    \includegraphics[width=0.4\textwidth]{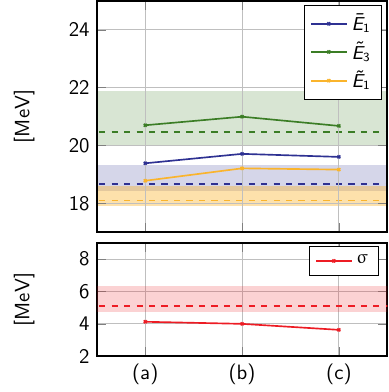}
    \caption{Mean energies (Eq.~\eqref{eq:weighted_av}) and dispersion (Eq.~\eqref{eq:stddev}) computed in $^{46}$Ti from the N$^{3}$LO ($\lambda_{\text{srg}}=1.88$~fm$^{-1}$) Hamiltonian ($e_{\!_{\;\text{max}}}=10, \hbar\omega=12$~MeV) and using the three sets of HFB states displayed in Fig.~\ref{fig:PES_2d_points} into the PGCM ansatz.}
    \label{fig:mom_2d_points}
\end{figure}

The HFB quadrupole deformation $\beta_2$, which is left free to adjust along the 1D TES, is displayed in the lower panel of Fig.~\ref{fig:1d_PES}. It is seen to vary almost linearly with $r$, a slight deviation from the linear trend being only detectable for small $\beta_2$ values (compression). Because of this feature of the 1D calculation, the energy of the GQR (not shown here) is strongly correlated with the GMR and is in fact located at the same energy. Again, the response is independent of the chosen mesh.

Eventually, one concludes that a reasonable choice of the mesh discretization step induces a negligible uncertainty in 1D PGCM calculations.



\paragraph{Energy interval}

The same 1D setting is now exploited to define four sets of HFB states according to the maximum excitation energy above the minimum, i.e. $E_{\text{max}}=$ 10, 20, 30 and 40~MeV using a fixed mesh value of $0.05$~fm. As seen in Fig.~\ref{fig:1d_PES_E}, it corresponds to performing PGCM calculations based on 9, 13, 16 and 18 HFB states, respectively. 

\begin{figure*}
    \centering
    \includegraphics[width=0.95\textwidth]{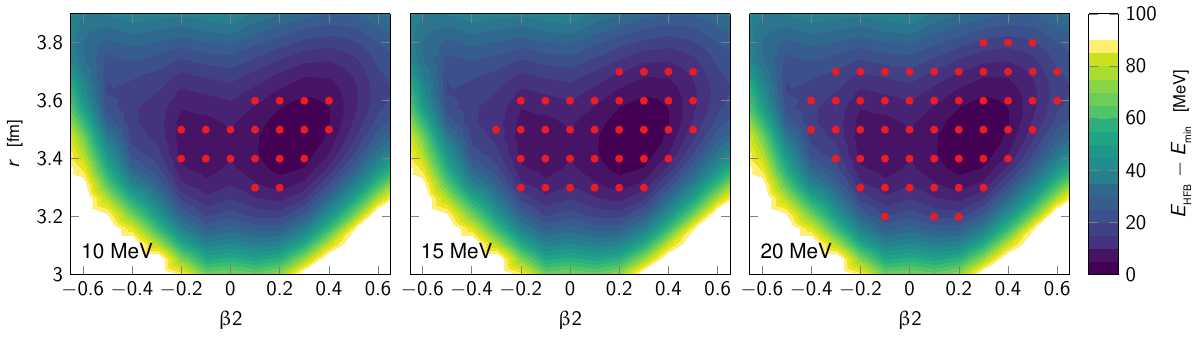}
    \caption{Same as Fig.~\ref{fig:PES_2d_points} but selecting HFB points according to their excitation energy above the minimum, i.e. from left to right the maximum energy is $E_{\text{max}}=10$~MeV (19 points), 15~MeV (34 points) and 20~MeV (51 points).}
    \label{fig:PES_2d_E}
\end{figure*}

\begin{figure}
    \centering
    \includegraphics[width=0.45\textwidth]{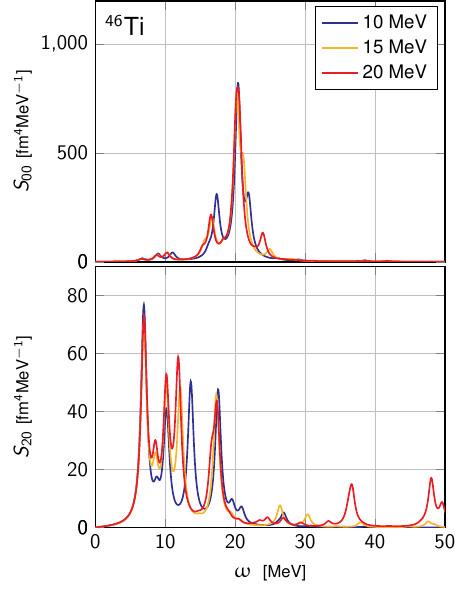}
    \caption{Same as Fig.~\ref{fig:spectra_2d_points} for HFB points with a maximum excitation energy above the minimum equal to $E_{\text{max}}=10$, 15 and 20~MeV.}
    \label{fig:spectra_2d_E}
\end{figure}

\begin{figure}
    \centering
    \includegraphics[width=0.4\textwidth]{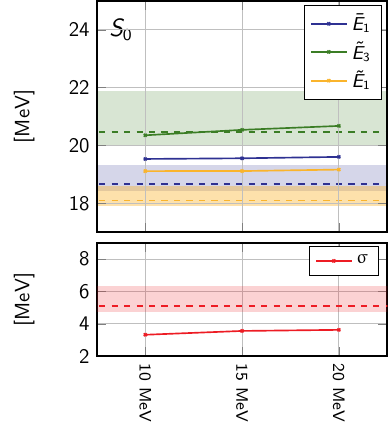}
    \caption{Same as Fig.~\ref{fig:mom_2d_points} for HFB points with a maximum excitation energy above the minimum equal to $E_{\text{max}}=10$, 15 and 20~MeV.}
    \label{fig:mom_2d_E}
\end{figure}

The resulting monopole responses are shown in Fig.~\ref{fig:1d_spectra_E}. In logarithmic scale, tiny variations are again only visible for high-energy states, the convergence being anyway quickly achieved there as well. The GMR containing a single peak located at $16.8$\,MeV in the 1D calculation is already fully converged for $E_{\text{max}}=10$~MeV. It is striking to observe that PGCM eigenstates located between 50 and 100~MeV excitations are fully converged using HFB seeds located only up to 40~MeV (or less) above the HFB minimum.



\subsubsection{Two-dimensional case}

Realistic PGCM calculations aiming at investigating the coupling between monopole and quadrupole modes require the use of both $r$ and $\beta_2$ as generator coordinates. Thus, the uncertainty associated with the selection of HFB states in the $(r,\beta_2)$ plane is now assessed via 2D PGCM calculations.

\paragraph{Mesh resolution}

Based on the 2D HFB TES obtained from the N$^{3}$LO ($\alpha=$0.08 fm$^4$) Hamiltonian ($\hbar\omega=$12~MeV and $\;e_{\!_{\;\text{max}}}=10$), three sets of HFB states are defined according to different grids in the $(r,\beta_2)$ plane. The canonical grid used so far corresponds to panel (c) in Fig.~\ref{fig:PES_2d_points}. Starting from there, selected points are rarefied by withdrawing one every two HFB states, in two successive steps. This results in the number of HFB points being reduced by a factor of two in panel (b) and four in panel (a).

\begin{figure*}
    \centering
    \includegraphics[width=1.02\textwidth]{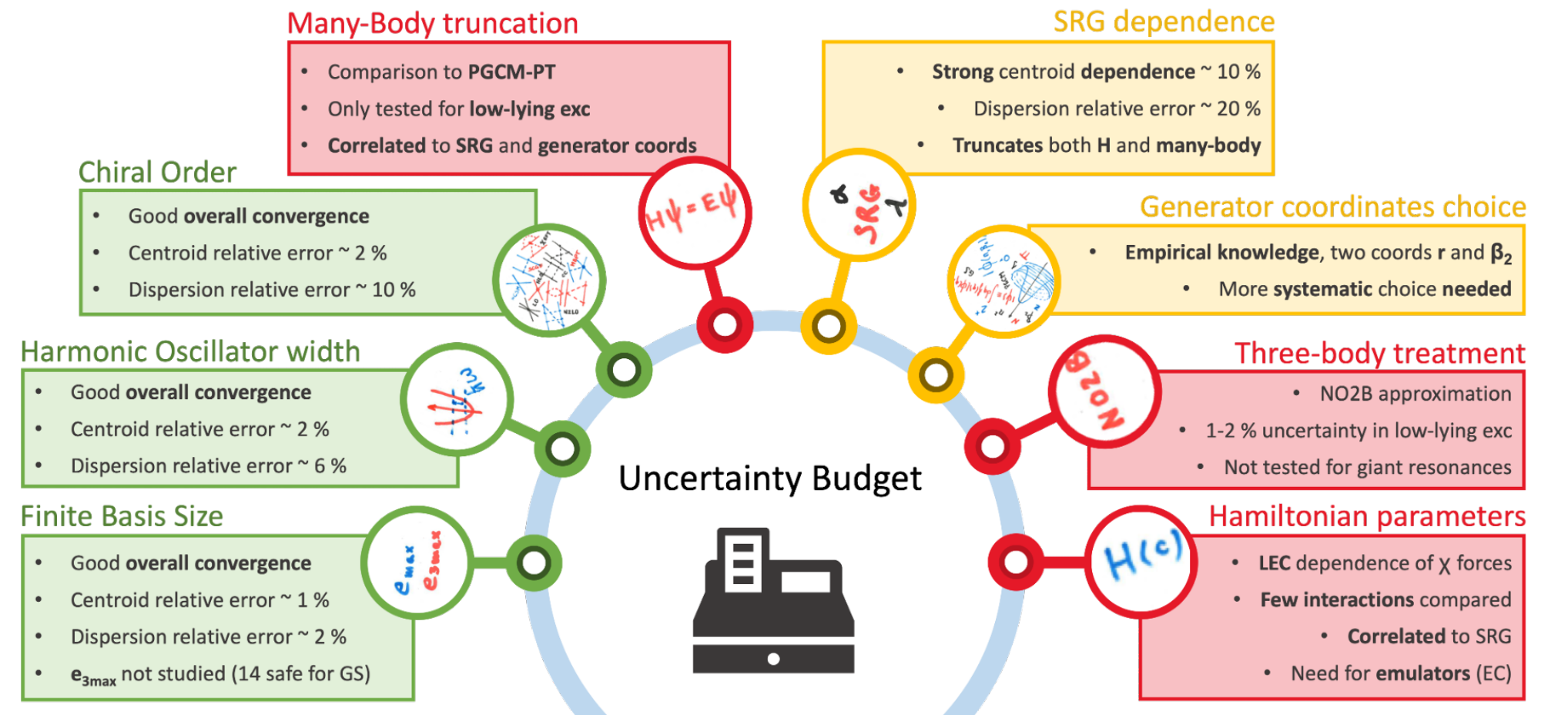}
    \caption{Summary of the uncertainty budget. In green are indicated the uncertainties that were thoroughly investigated. In yellow are those that could only be touched upon. Eventually, boxes in red correspond to those that could at best be estimated from previous but somewhat different works or not estimated at all.}
    \label{fig:enter-label}
\end{figure*}

The monopole and quadrupole responses are displayed in Fig.~\ref{fig:spectra_2d_points}. The energy of the main GMR peak is very stable near $20.5$~MeV. On the other hand, secondary peaks only develop when using denser meshes while the strength of the main peak changes accordingly. In particular, the secondary peak at $18.6$~MeV corresponding to the GMR in the previous 1D calculation is absent with mesh (a) and not yet converged to the right value with mesh (b). Looking at the quadrupole response, one observes that  mesh (c) is necessary to converge the overall strength function and locate the peak at $18.6$~MeV at the right position\footnote{This peak is the only one appearing in 1D calculations, both in the monopole and in the quadrupole response. While these correlated peaks survive in the 2D calculation, this does not correspond to the main GMR peak in the monopole channel. As discussed at length in Paper II, this secondary peak results from the coupling between the monopole and quadrupole modes whenever the nucleus is intrinsically deformed.}.

The mean energies and dispersion of the monopole strength function are shown in Fig.~\ref{fig:mom_2d_points}. They are very stable with respect to the increase of the mesh density, i.e. the centroid energy (dispersion) is estimated to be converged to better than $0.5\%$ ($2\%$).  


\paragraph{Energy window}

The uncertainty of the PGCM results with respect to the maximum energy $E_{\text{max}}$ of the selected HFB states is now evaluated. To do so, the three values $E_{\text{max}}=10$, 15 and 20~MeV are utilised, as illustrated in Fig.~\ref{fig:PES_2d_E}. They correspond to performing 2D PGCM calculations with 19 , 34 and 51 HFB states, respectively.

The monopole and quadrupole responses are displayed in Fig.~\ref{fig:spectra_2d_E}. In the monopole channel, all important features are already well converged for $E_{\text{max}}=15$, including the main GMR peak and the peak associated with the coupling to the GQR at $16.5$~MeV. 

The situation is unusually satisfying for the quadrupole response. While for $E_{\text{max}}=10$~MeV the GQR is already converged, structures at lower energies are well described for $E_{\text{max}}=15$~MeV. One however notes that peaks beyond $35$\,MeV emerge for $E_{\text{max}}=20$~MeV and are probably not converged yet.

The mean energies and  dispersion of the monopole strength function are shown in Fig.~\ref{fig:mom_2d_E} as a function of $E_{\text{max}}$. They are very stable and carry uncertainties below $0.5\%$. 



\section{Summary and conclusions}
\label{sec:concl}

A systematic analysis of ab initio calculations of the isoscalar monopole strength function via the PGCM has been presented for the nucleus $^{46}$Ti.
This systems is considered to be representative of light- and medium-mass (doubly open-shell) nuclei, for which similar conclusions are thus expected to apply.
The results of our investigations are summarised in Fig.~\ref{fig:enter-label} giving a compact view of the outcome of the present study. 
For the three uncertainties that could be thoroughly assessed, a good stability of the PGCM calculations of the monopole response could be demonstrated. 
More specifically, uncertainties on the centroid energy were shown to remain below $2\%$ whereas the dispersion is uncertain on the $10\%$ level. 

Clearly, much remains to be done to achieve a full control on the uncertainties of \textit{ab initio} PGCM calculations of GRs. Still, the intermediate conclusion that can be presently reached is that such calculations are pertinent and can be quantitative, at least regarding peaks carrying a large fraction of the strength. The following Paper II builds on such a conclusion to present a series of results of physical and experimental interest. Eventually, the mean-field-like scaling with system's size of the PGCM computational cost makes the method an excellent candidate to extend  \textit{ab initio} studies of collective excitations to yet heavier closed- and open-shell nuclei in the future.

\section*{Acknowledgements}

Calculations were performed by using HPC resources from GENCI-TGCC (Contract No. A0130513012). A.P. was supported by the CEA NUMERICS program, which has received funding from the European Union's Horizon 2020 research and innovation program under the Marie Sk{\l}odowska-Curie grant agreement No 800945, and by the Deutsche Forschungsgemeinschaft (DFG, German Research Foundation) – Projektnummer 279384907 – SFB 1245.

\section*{Data Availability Statement}

This manuscript has no associated data or the data will not be deposited.

\bibliography{biblio.bib}

\end{document}